\documentclass[amstex,epsf,amssymb,usenatbib]{mn2e}
\usepackage{epsf}
\usepackage{amsmath,amssymb,graphicx}
\def\araa{ARA\&A}             % Annual Review of Astron and Astrophys
\def\apj{ApJ}                 % Astrophysical Journal
\def\apjl{ApJ}                % Astrophysical Journal, Letters
\def\aap{A\&A}                % Astronomy and Astrophysics
\def\mnras{MNRAS}             % Monthly Notices of the RAS
\title{The influence of the Mach number on the stability of radiative shocks}

\author[B. Ramachandran \& M.D. Smith]
       {Babulakshmanan Ramachandran$^{1}$\thanks{E-mail: brc@arm.ac.uk}
\& Michael D. Smith $^{1,2}$\thanks{E-mail: m.d.smith@kent.ac.uk} \\
$^1$Armagh Observatory, College Hill, Armagh BT61 9DG, Northern Ireland, U.K.\\
$^2$Centre for Astrophysics \& Planetary Science, The University of Kent, Canterbury, Kent
CT2 7NR, U.K. }

\date{Accepted .....
      Received ..... ;
      in original form .....}

\pagerange{\pageref{firstpage}--\pageref{lastpage}}
\pubyear{2005}

\begin{document}

\maketitle

\label{firstpage}

\begin{abstract}  
\noindent
We study the stability properties of  hydrodynamic shocks with finite Mach numbers.
The linear analysis supplements previous analyses which took the strong shock limit. 
We derive the linearised equations for a general specific heat ratio as well as 
temperature and density power-law cooling functions, corresponding to a range of conditions relevant 
to interstellar atomic and molecular cooling processes. Boundary conditions corresponding
to a return to the upstream temperature ($R$\,=\,1) and to a cold wall ($R$\,=\,0) are investigated.
We find that for Mach number $M > 5$, the strong shock overstability limits are not 
significantly modified. For  $M < 3$, however, shocks are considerably more stable for most cases. 
In general, as the shock weakens, the critical values of the temperature power-law index (below
which shocks are overstable)
are reduced for the overtones more than 
for the fundamental, which signifies a change in basic behaviour. In the $R$\,=\,0 scenario, 
however, we find that the overstability regime and growth rate of the fundamental mode are increased 
when cooling is under local thermodynamic equilibrium. We provide a possible explanation for the 
results in terms of a 
stabilising influence provided downstream but a destabilising effect 
associated with the shock front. We conclude that the regime of overstability
for interstellar atomic shocks is well represented 
by the strong shock limit unless the upstream gas is hot. Although molecular shocks can be overstable
to overtones, the magnetic field provides a significant stabilising influence.
\end{abstract}


%%%%%%%%%%%%%%%%%%%%%%%%%%%%%%%%%%%%%%%%%%%%%%%%%%%%%%%%%%%%%%%%%%%%%%%%
\section{Introduction}              %%%%%%%%%%%%%%%%%%  INTRO %%%%%%%%%%
%%%%%%%%%%%%%%%%%%%%%%%%%%%%%%%%%%%%%%%%%%%%%%%%%%
\label{intro}

Across a shock front, a fraction of the bulk kinetic energy is thermalised. 
Across interstellar shock waves, the shock front is followed by a cooling layer
in which a fraction of the thermal energy escapes as radiation. Across a {\em radiative} shock wave,
the cooling time is much shorter than the dynamical evolution of the system. 
Such shocks propagate through many astrophysical media being driven by, for example, explosions, winds, jets
and collisions \citep{1987ip...symp..283S,1993ARA&A..31..373D}. 
The fraction of energy which is thermalised in a steady-state hydrodynamic shock front
can be simply expressed in terms of the specific heat ratio, $\gamma$ and the Mach number, $M$, of the 
upstream (pre-shock) flow \citep{1987ip...symp..283S}. However, in some circumstances, radiative shocks are prone 
to an overstability due to the nature of the cooling \citep[][hereafter CI82]{1982ApJ...261..543C}. In one-dimensional
simulations, the oscillations of growing amplitude can lead to a quasi-periodic collapse and 
restoration of almost the entire shock layer \citep[e.g.][]{2003ApJ...591..238S}. 

%%%%%%%%%%%%%%%%%%%%%%%%%%%%%% SKETCH %%%%%%%%%%%%%%%%%%%%%%%%%%%%%%%%%%%%%%%
\begin{figure}
\begin{center} 
\epsfxsize=8.7cm 
\epsfbox{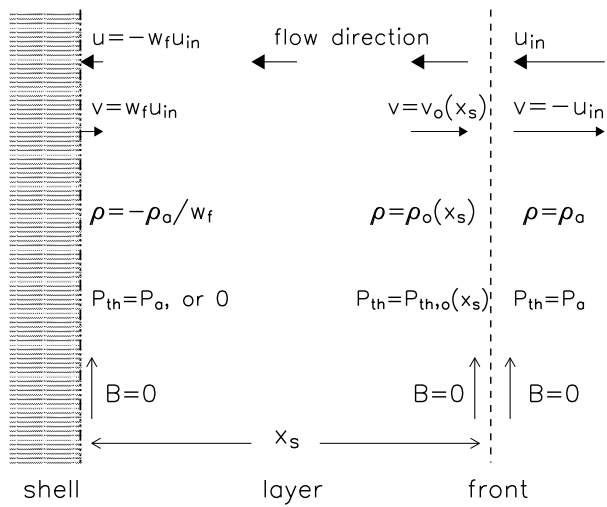}
\caption[]{A sketch of the steady weak shock configuration. The shock jump conditions are 
defined by Eqs~\ref{rhozero}, \ref{velzero} and \ref{pzero}. For weak shocks, we consider 
two cases: a final temperatures, $T_a$, corresponding to a shell of constant temperature 
equal to the upstream temperature or equal to zero, corresponding to accumulation onto a 
stationary wall.}
\label{steady}
\end{center}
\end{figure}
%%%%%%%%%%%%%%%%%%%%%%%%%%%%%%%%%%%%%%%%%%%%%%%%%%%%%%%%%%%%%%%%%%%%%%%%%%%%% 

The overstability was first discovered by \citet{1981ApJ...245L..23L}. Soon after,
\citet{1982ApJ...261..543C} performed a linear stability analysis for plane 
parallel flows with $\Lambda \propto \rho^{2}~T^{\alpha}$.
They found that such shocks were linearly unstable in a 
fundamental mode if the exponent $\alpha \lesssim 0.4$ and unstable to overtones for  
$\alpha \lesssim 0.8$. The physical basis for the instability is that, while 
the shock wave is moving away from the surface, it heats the gas to a higher 
temperature and the cooling time is longer than in the steady state case. As a
result, in the unstable regime, the shock structure attempts to form
an even larger cooling region. In contrast, while the shock wave is moving in, the 
situation is reversed thus yielding amplified oscillations. 

A wide range of numerical and linear analyses have now catered for hot atomic gases in various physical and geometric forms 
\citep[e.g.][]{1984ApJ...276..667I,1986ApJ...304..154B,1993ApJ...413..176T,1994ApJ...434..262D,1996ApJ...458..327I,1995ApJ...449..727S,2003ApJ...591..238S}. 
Recently, \citet[][Paper I]{2005MNRAS.362.1353R} provided a review while extending the linear analysis to 
show that the instability is not only sensitive to the cooling function but also to the specific heat ratio. 
\citet[][Paper II]{2005MNRAS.357..707R} analysed the corresponding magnetohydrodynamic case, thus considering
both molecular and atomic shock waves.
In fact, these linear analyses are only applicable to strong shocks in which the Mach number is assumed 
to be sufficiently large so that the shock profile is independent of the Mach number. Then, the
front compression is $(\gamma+1)/(\gamma-1)$ and the entire compression is
effectively infinite unless limited by the magnetic field. 

This leaves open the question of the dependence on the Mach number: for a given power-law cooling function, 
when is there a critical Mach number above which a shock is unstable? In fact,
\citet{1995ApJ...449..727S} performed numerical simulations taking $\gamma = 5/3$ and found that 
low Mach number shocks were more stable than shocks with high Mach number. At lower $M$, it was 
found that the lower density in the cold gas layer acts like a shock absorber. In contrast, at 
high $M$, the layer is like a reflecting wall that rebounds incident waves.
Although not comprehensive, a set of displayed simulations indicated that a shock unstable to overtone 
modes at high Mach number became stable for $M$ somewhere in the range 5\,--\,10.
Recently, \citet{2005A&A...438...11P} have also studied
weak shocks numerically for a variety of boundary conditions and find that Mach numbers of order 100
are required before they are classified as strong shocks. They also find that the stability increases
with the decrease in Mach number. Their study reveals that if the lower boundary condition is such that 
the post-shock gas cools to a temperature below the pre-shocked value, low Mach
number shocks reach critical values of $\alpha$ which may be comparable to high Mach number shocks under 
the boundary condition in which the pre-shock and post-shock temperatures are the same.

To complement the above investigations as well as to extend the problem to molecular shocks, we have 
undertaken here a one-dimensional linear stability analysis with the Mach number as the main parameter while  
varying the specific heat ratio and two parameters, $\alpha$ and $\beta$, which describe the cooling per unit volume
in the form $\Lambda \propto \rho^{\beta}T^{\alpha}$. For evaluation, we pose a second question: for 
a given $\beta$ and $\gamma$, how low must the Mach number be in order to significantly decrease the 
critical value of $\alpha$ from the strong shock value?

%%%%%%%%%%%%%%%%%%%Figures for deltar versus alpha %%%%%%%%%%%%%%%%%%%%%%%%%%%
%%%%%%%%%%%%%%%%%%%%%%%%%%%%%%Figure 2%%%%%%%%%%%%%%%%%%%%%%%%%%%%%%%%%%%%%%%
\begin{figure*}
\centering   
	\includegraphics[width=15.2cm]{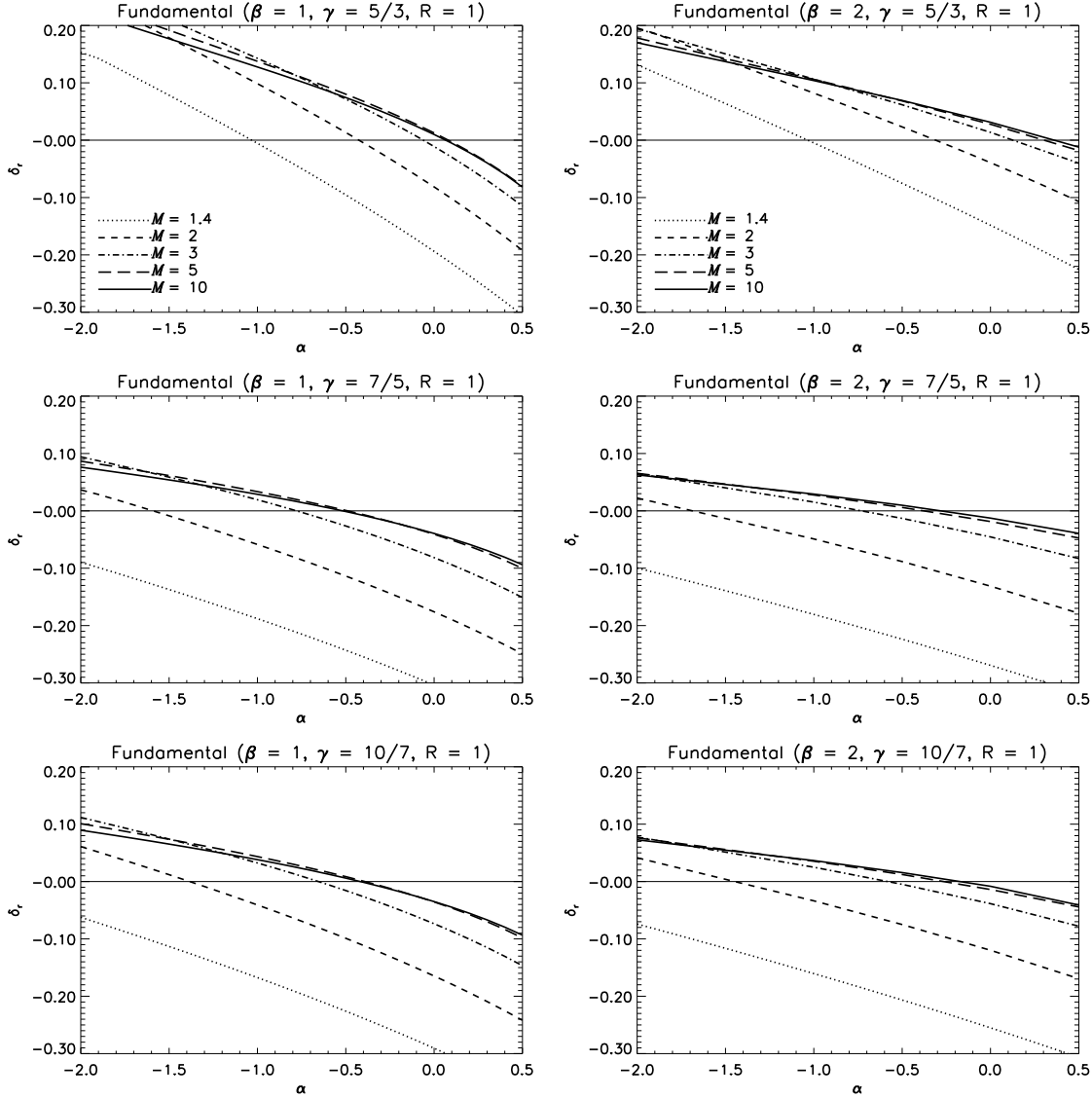} 
\caption{The growth/damping rates, $\delta_{r}$, as a function of $\alpha$ for the 
fundamental mode with R\,=\,1. Values of $\gamma$, $\beta$  and $M$ are indicated.}
\label{fundrR1}
\end{figure*} 
%%%%%%%%%%%%%%%%%%%%%%%%%%%%%%Figure 2%%%%%%%%%%%%%%%%%%%%%%%%%%%%%%%%%%%%%%%
%%%%%%%%%%%%%%%%%%%%%%%%%%%%%%Figure 3%%%%%%%%%%%%%%%%%%%%%%%%%%%%%%%%%%%%%%%
\begin{figure*}
\centering   
	\includegraphics[width=15.2cm]{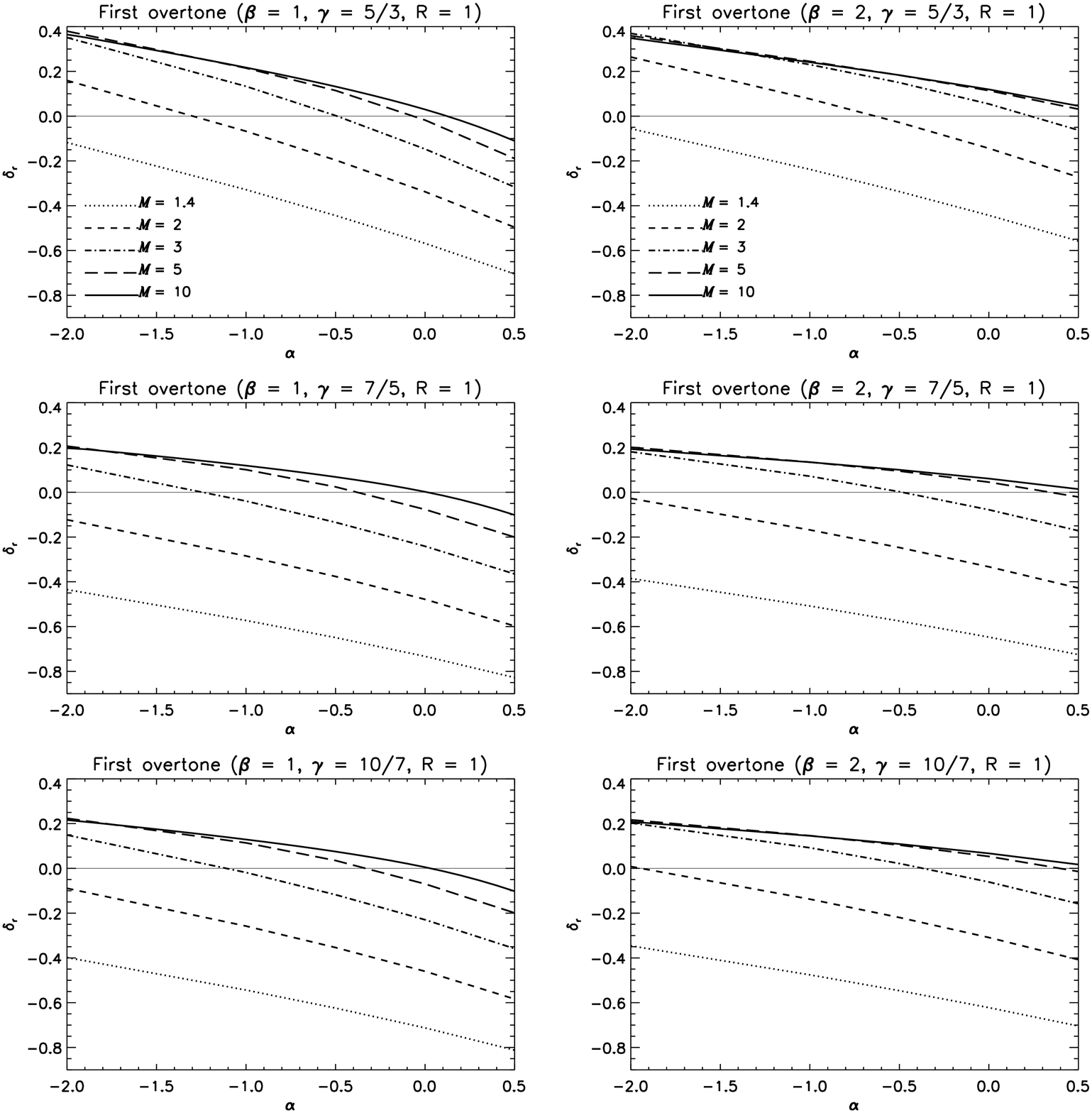} 
\caption{The growth/damping rates, $\delta_{r}$, as a function of $\alpha$ for the 
first overtone mode with R\,=\,1. Values of $\gamma$, $\beta$  and $M$ are indicated.}
\label{firstrR1}
\end{figure*} 
%%%%%%%%%%%%%%%%%%%%%%%%%%%%%%Figure 3%%%%%%%%%%%%%%%%%%%%%%%%%%%%%%%%%%%%%%
%%%%%%%%%%%%%%%%%%%%%%%%%%%%%%Figure 4%%%%%%%%%%%%%%%%%%%%%%%%%%%%%%%%%%%%%%%
\begin{figure*}
\centering   
	\includegraphics[width=15.2cm]{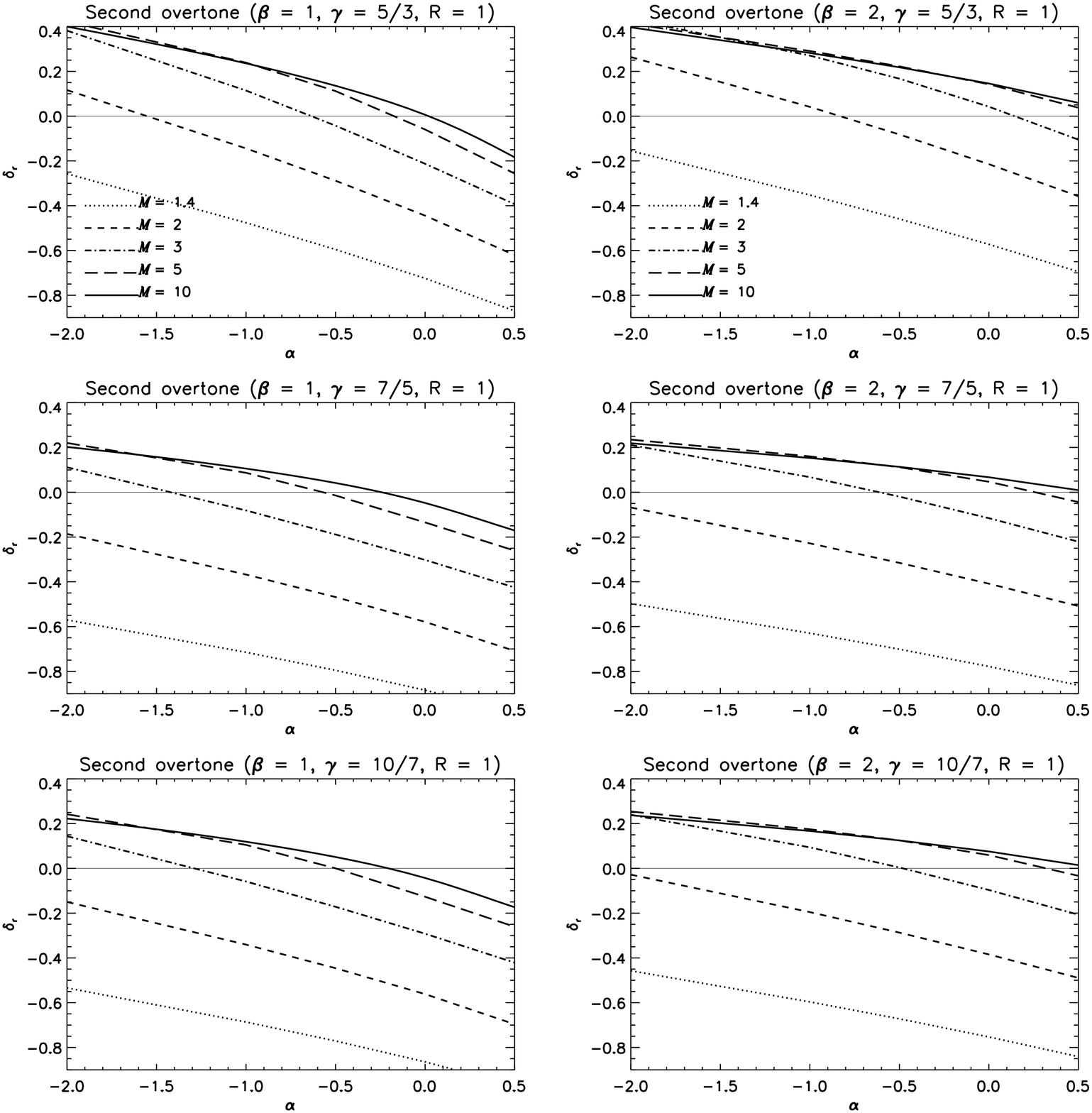} 
\caption{The growth/damping rates, $\delta_{r}$, as a function of $\alpha$ for the 
second overtone with R\,=\,1. Values of $\gamma$, $\beta$  and $M$ are indicated.}
\label{secrR1}
\end{figure*} 
%%%%%%%%%%%%%%%%%%%%%%%%%%%%%%Figure 4%%%%%%%%%%%%%%%%%%%%%%%%%%%%%%%%%%%%%%%
%%%%%%%%%%%%%%%%%%%%%%%%%%%%%%Figure 5%%%%%%%%%%%%%%%%%%%%%%%%%%%%%%%%%%%%%%%
\begin{figure*}
\centering   
	\includegraphics[width=15.2cm]{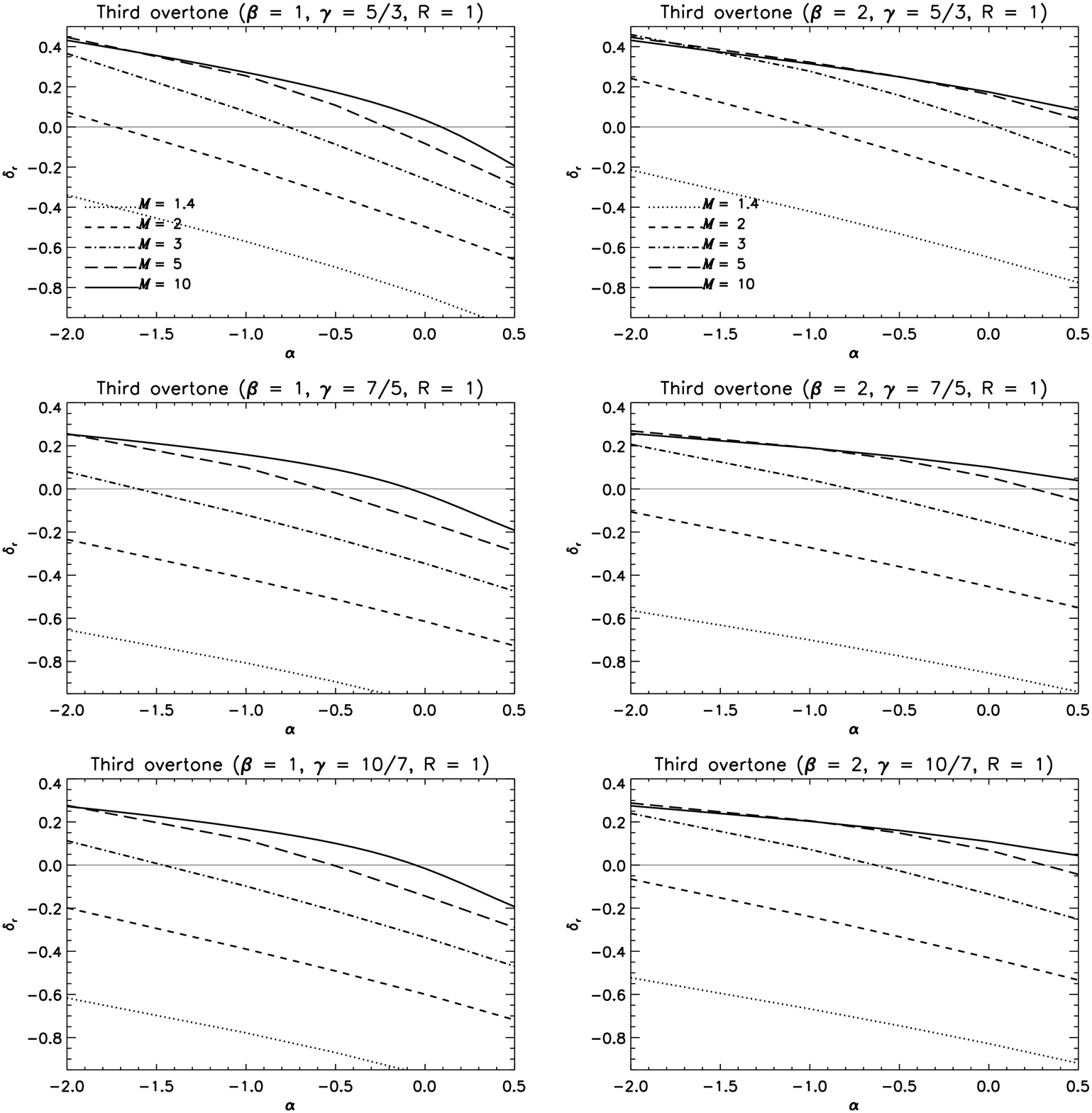}
\caption{The growth/damping rates, $\delta_{r}$, as a function of $\alpha$ for the 
third overtone with R\,=\,1. Values of $\gamma$, $\beta$  and $M$ are indicated.} 
\label{thirdrR1}
\end{figure*} 
%%%%%%%%%%%%%%%%%%%%%%%%%%%%%%Figure 5%%%%%%%%%%%%%%%%%%%%%%%%%%%%%%%%%%%%%%%

%%%%%%%%%%%%%%%%%%%%%%%%%% R = 0 begins %%%%%%%%%%%%%%%%%%%%%%%%%%%%%%%%%%%
%%%%%%%%%%%%%%%%%%%%%%%%%%%%%%Figure 6%%%%%%%%%%%%%%%%%%%%%%%%%%%%%%%%%%%%%%%
\begin{figure*}
\centering   
	\includegraphics[width=15.2cm]{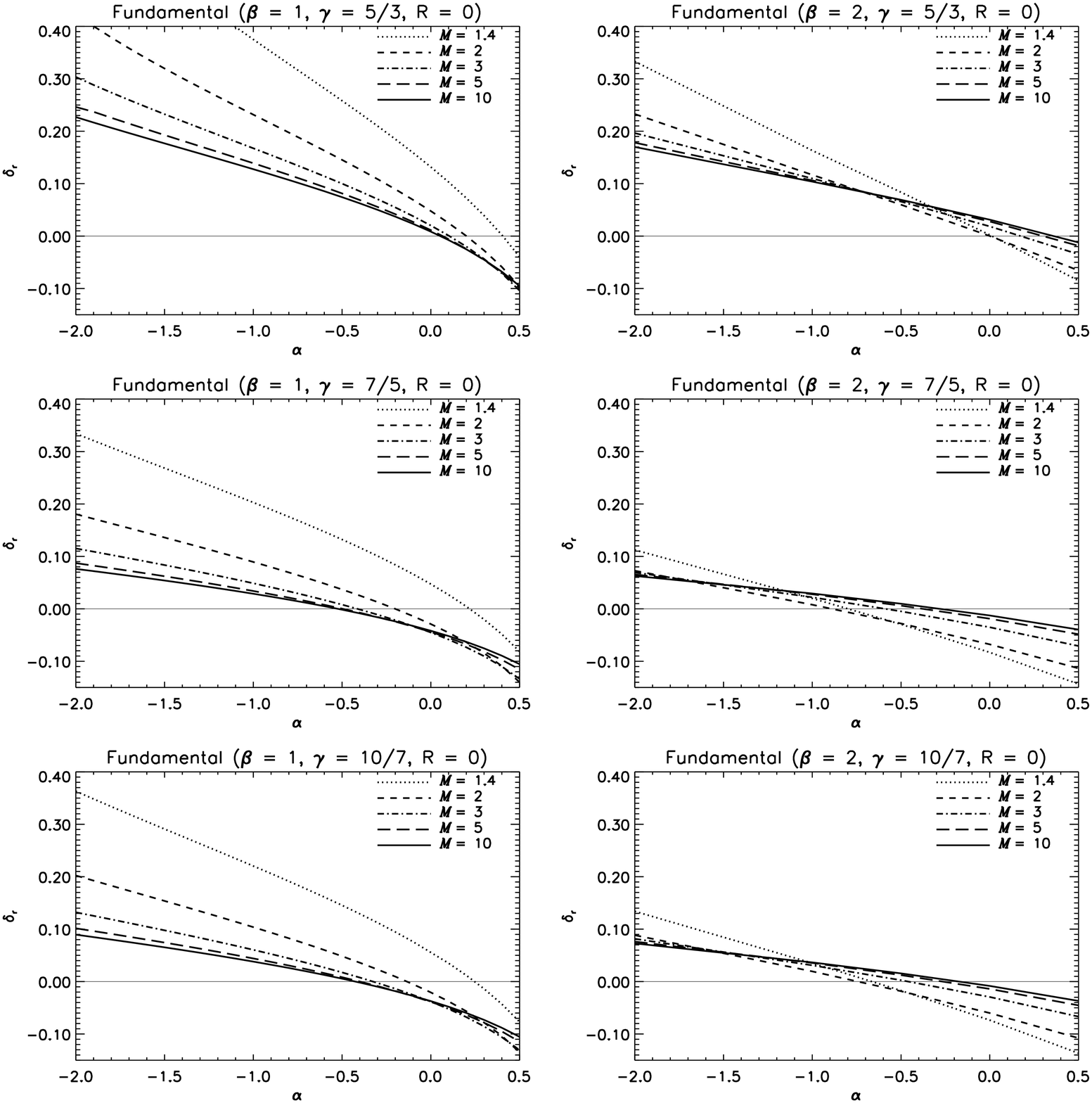}
\caption{The growth/damping rates, $\delta_{r}$, as a function of $\alpha$ for the 
fundamental mode with R\,=\,0. Values of $\gamma$, $\beta$  and $M$ are indicated.} 
\label{fundrR0}
\end{figure*} 
%%%%%%%%%%%%%%%%%%%%%%%%%%%%%%Figure 6%%%%%%%%%%%%%%%%%%%%%%%%%%%%%%%%%%%%%%%
%%%%%%%%%%%%%%%%%%%%%%%%%%%%%%Figure 7%%%%%%%%%%%%%%%%%%%%%%%%%%%%%%%%%%%%%%%
\begin{figure*}
\centering   
	\includegraphics[width=15.2cm]{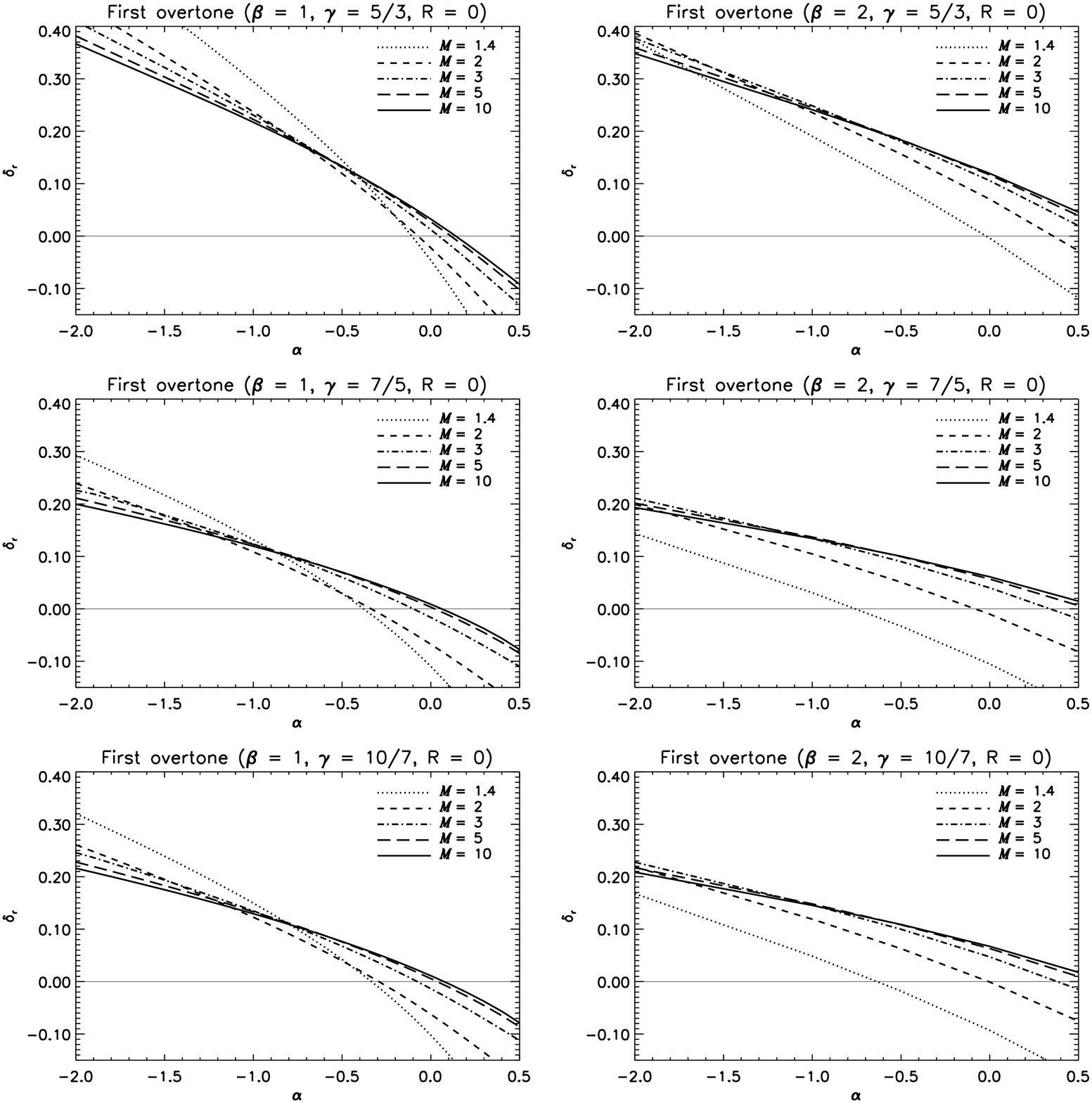}
\caption{The growth/damping rates, $\delta_{r}$, as a function of $\alpha$ for the 
first overtone mode with R\,=\,0. Values of $\gamma$, $\beta$  and $M$ are indicated.} 
\label{firstrR0}
\end{figure*} 
%%%%%%%%%%%%%%%%%%%%%%%%%%%%%%Figure 7%%%%%%%%%%%%%%%%%%%%%%%%%%%%%%%%%%%%%%%
%%%%%%%%%%%%%%%%%%%%%%%%%%%%%%Figure 8%%%%%%%%%%%%%%%%%%%%%%%%%%%%%%%%%%%%%%%
\begin{figure*}
\centering   
	\includegraphics[width=15.2cm]{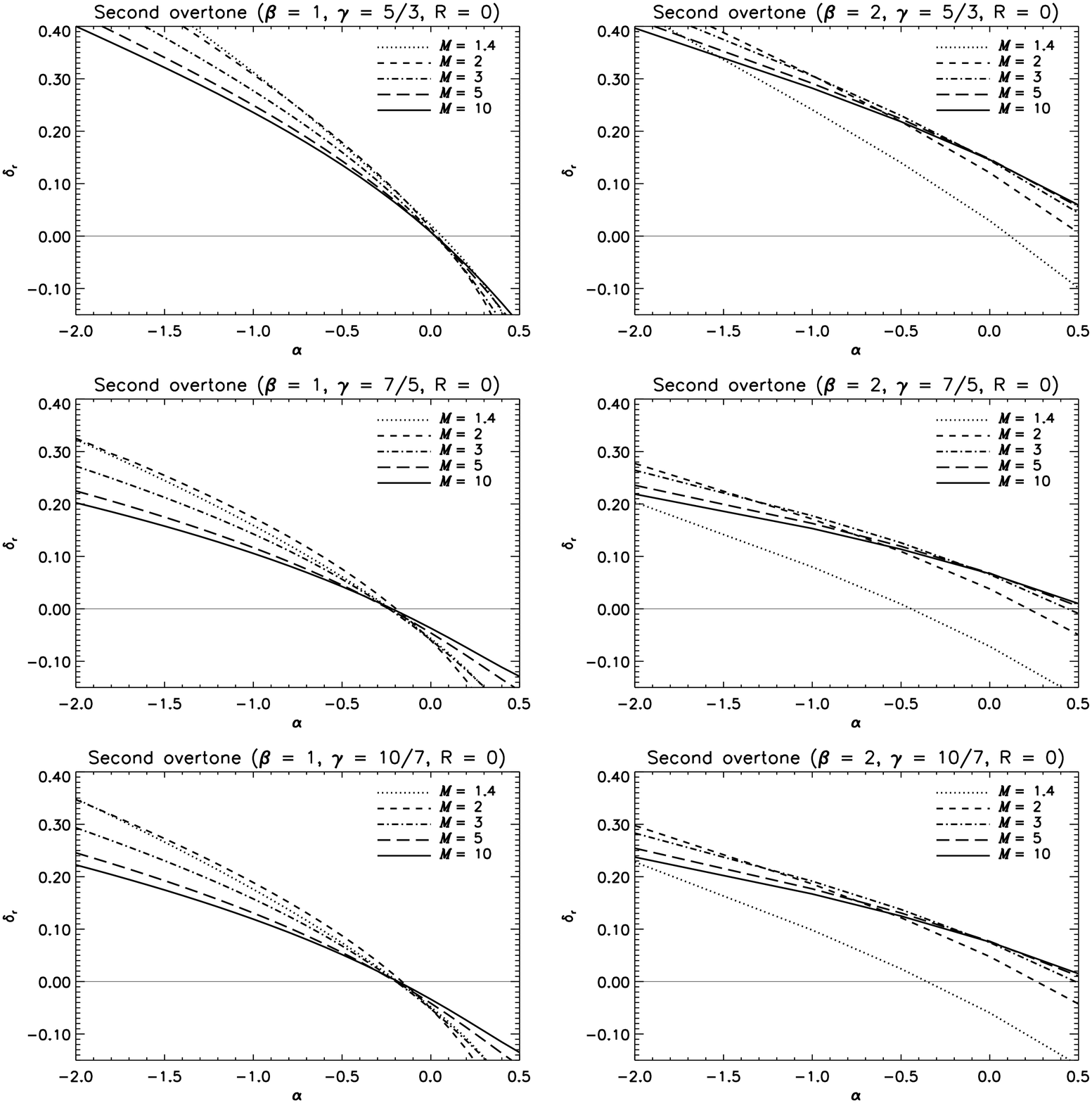}
\caption{The growth/damping rates, $\delta_{r}$, as a function of $\alpha$ for the 
second overtone with R\,=\,0. Values of $\gamma$, $\beta$  and $M$ are indicated.}  
\label{secrR0}
\end{figure*} 
%%%%%%%%%%%%%%%%%%%%%%%%%%%%%%Figure 8%%%%%%%%%%%%%%%%%%%%%%%%%%%%%%%%%%%%%%%
%%%%%%%%%%%%%%%%%%%%%%%%%%%%%%Figure 9%%%%%%%%%%%%%%%%%%%%%%%%%%%%%%%%%%%%%%%
\begin{figure*}
\centering   
	\includegraphics[width=15.2cm]{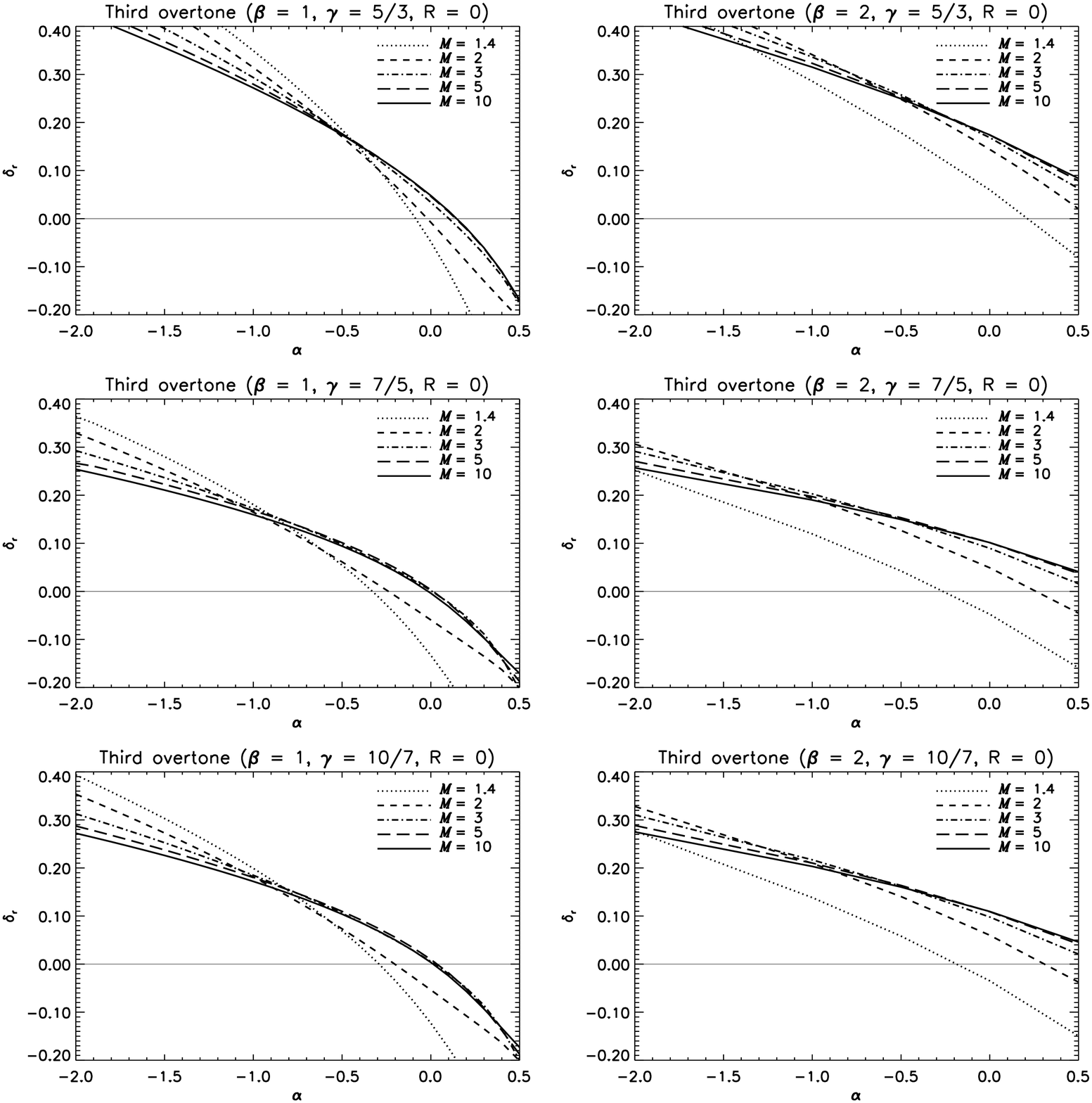}
\caption{The growth/damping rates, $\delta_{r}$, as a function of $\alpha$ for the 
third overtone with R\,=\,0. Values of $\gamma$, $\beta$  and $M$ are indicated.}   
\label{thirdrR0}
\end{figure*} 
%%%%%%%%%%%%%%%%%%%%%%%%%%%%%%Figure 9%%%%%%%%%%%%%%%%%%%%%%%%%%%%%%%%%%%%%%%

%%%%%%%%%%%%%%%%%%%%%%%%%%%%%%%%%%%%%%%%%%%%%%%%%%%%%%%%%%%%%%%%%%%%%%%%%%%%%%

%%%%%%%%%%%%%%%%%%%Figures for deltai versus alpha %%%%%%%%%%%%%%%%%%%%%%%%%%%
%%%%%%%%%%%%%%%%%%%%%%%%%%%%%%Figure 10%%%%%%%%%%%%%%%%%%%%%%%%%%%%%%%%%%%%%%%
\begin{figure*}
\centering   
	\includegraphics[width=15.2cm]{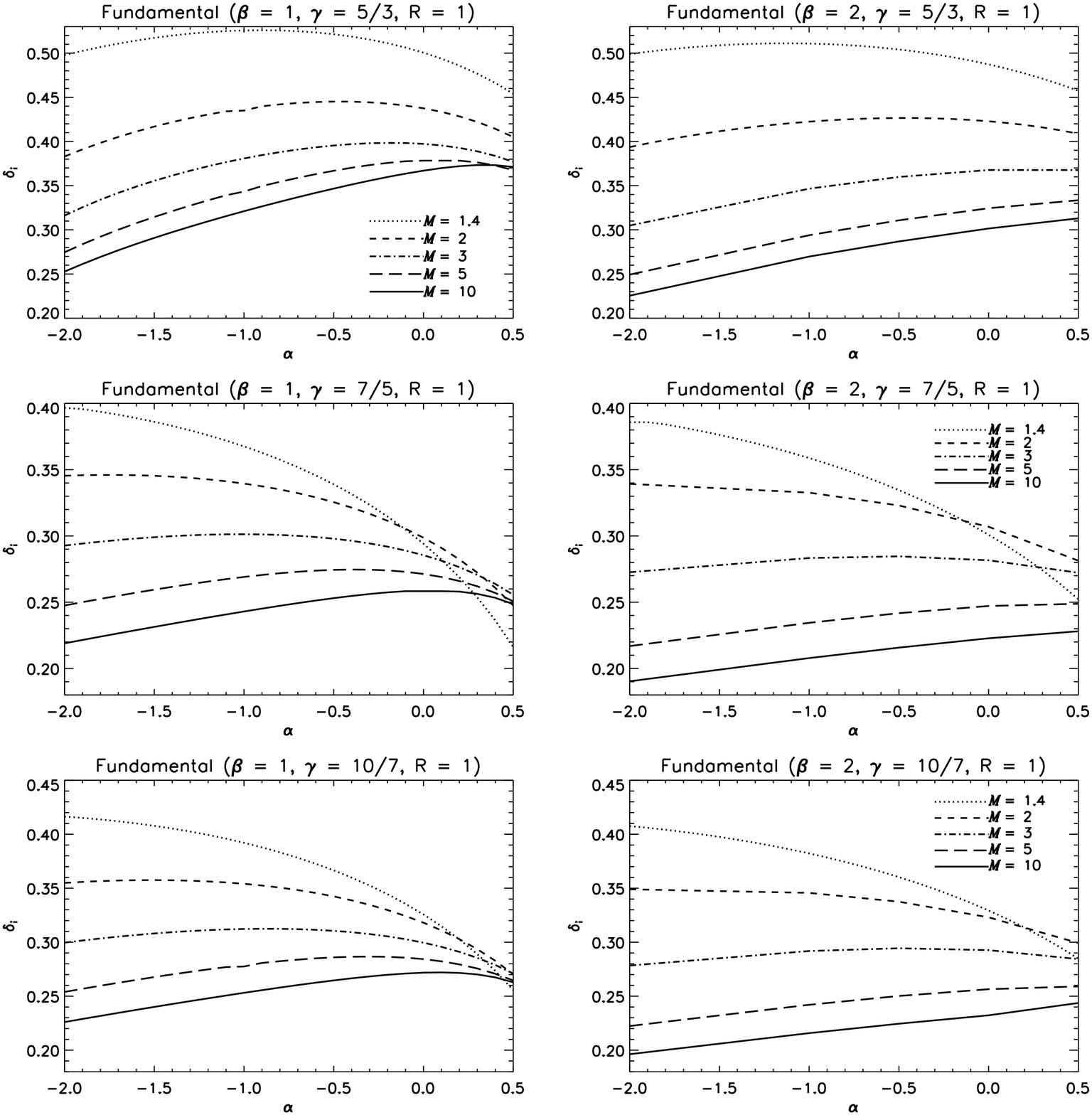} 
\caption{ The eigenfrequency, $\delta_{i}$ as a function of $\alpha$ for the 
fundamental mode  with R\,=\,1. Values of $\gamma$, $\beta$  and $M$ are indicated.}
\label{fundiR1}
\end{figure*} 
%%%%%%%%%%%%%%%%%%%%%%%%%%%%%%Figure 10%%%%%%%%%%%%%%%%%%%%%%%%%%%%%%%%%%%%%%%
%%%%%%%%%%%%%%%%%%%%%%%%%%%%%%Figure 11%%%%%%%%%%%%%%%%%%%%%%%%%%%%%%%%%%%%%%%
\begin{figure*}
\centering   
	\includegraphics[width=15.2cm]{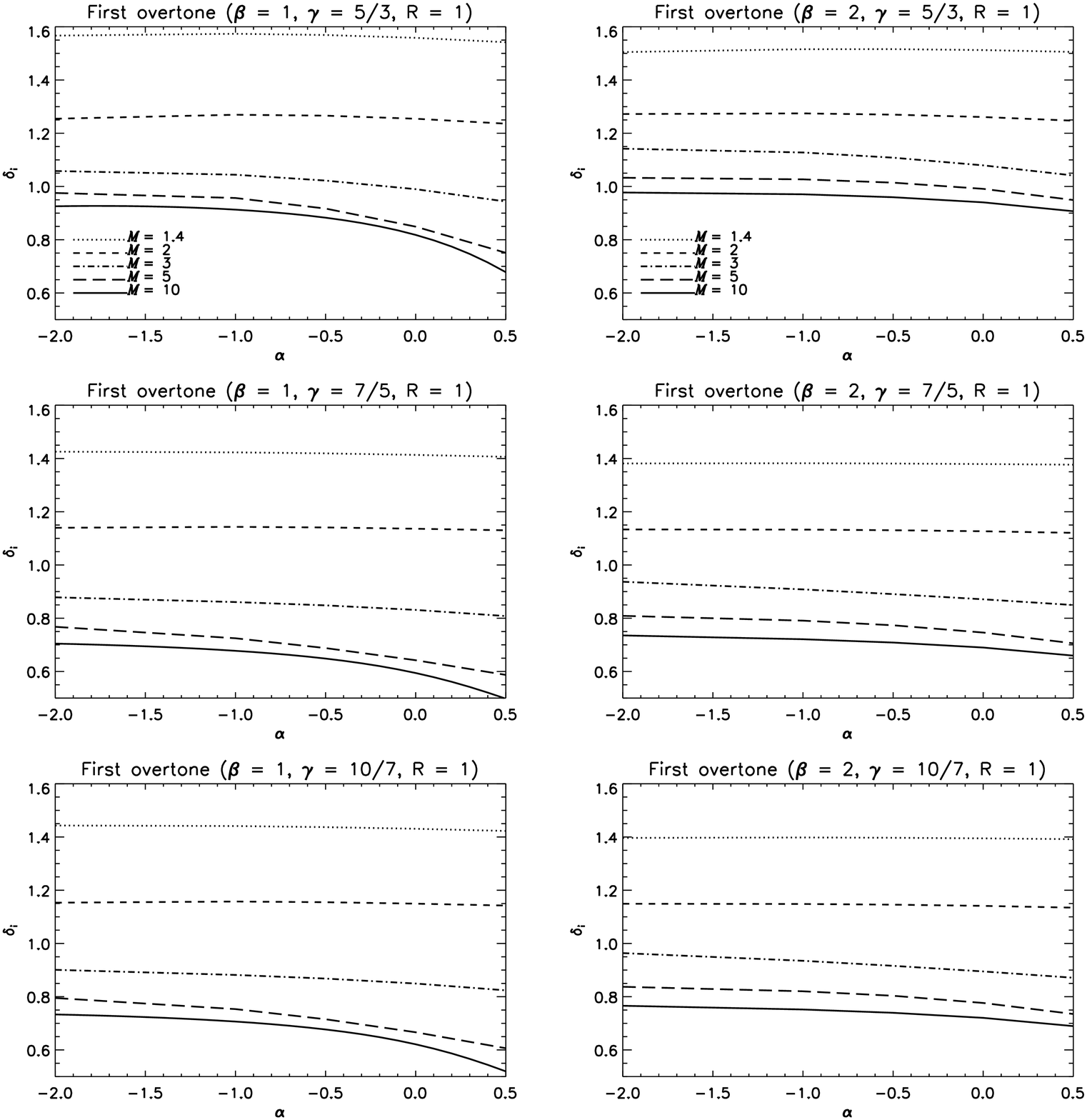} 
\caption{ The eigenfrequency, $\delta_{i}$ as a function of $\alpha$ for the 
first overtone mode  with R\,=\,1. Values of $\gamma$, $\beta$  and $M$ are indicated.}
\label{firstiR1}
\end{figure*} 
%%%%%%%%%%%%%%%%%%%%%%%%%%%%%%Figure 11%%%%%%%%%%%%%%%%%%%%%%%%%%%%%%%%%%%%%%
%%%%%%%%%%%%%%%%%%%%%%%%%%%%%%Figure 12%%%%%%%%%%%%%%%%%%%%%%%%%%%%%%%%%%%%%%%
\begin{figure*}
\centering   
	\includegraphics[width=15.2cm]{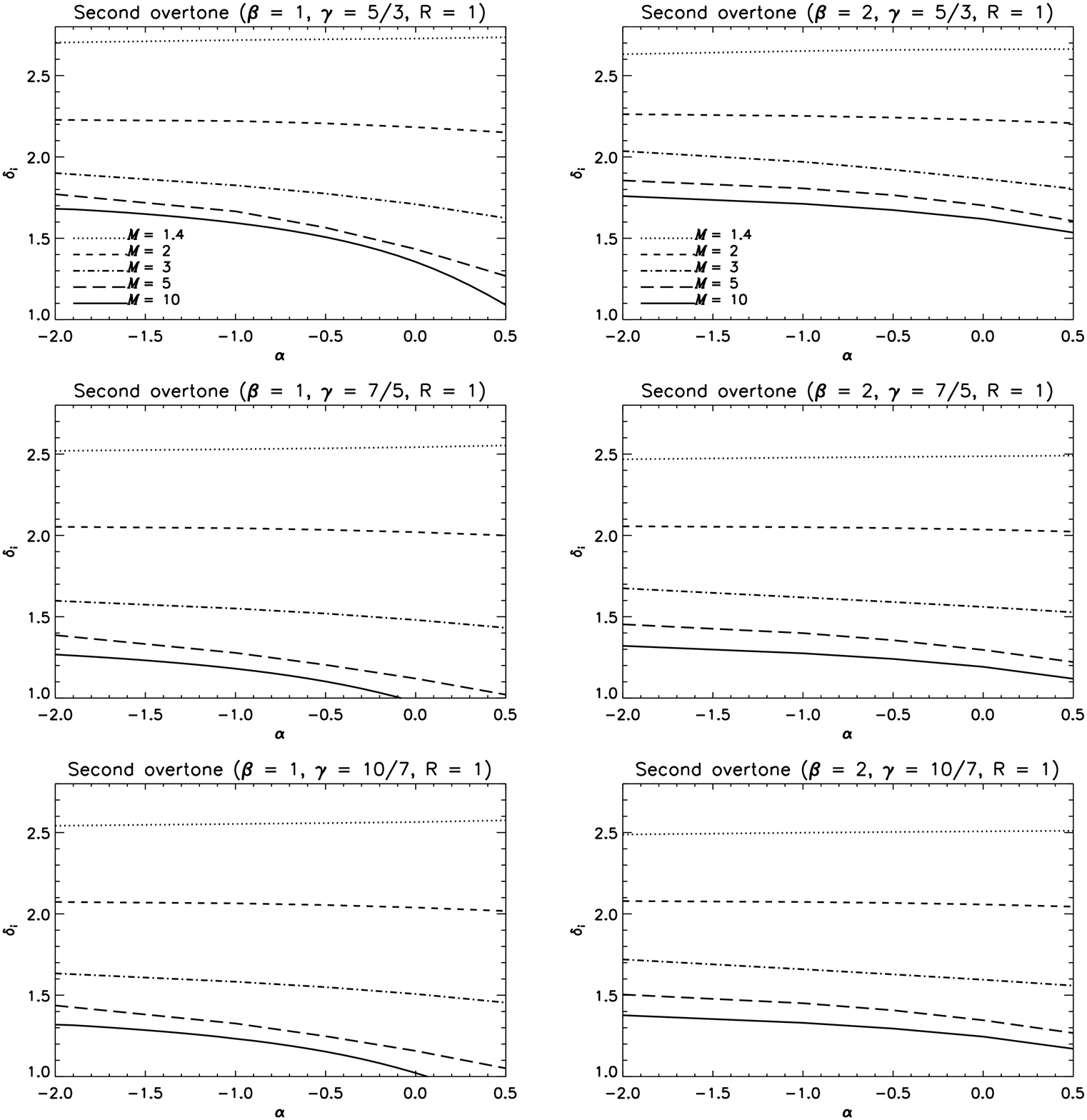} 
\caption{ The eigenfrequency, $\delta_{i}$ as a function of $\alpha$ for the 
second overtone mode  with R\,=\,1. Values of $\gamma$, $\beta$  and $M$ are indicated.}
\label{seciR1}
\end{figure*} 
%%%%%%%%%%%%%%%%%%%%%%%%%%%%%%Figure 12%%%%%%%%%%%%%%%%%%%%%%%%%%%%%%%%%%%%%%%

%%%%%%%%%%%%%%%%%%%%%%%%%%%%%%Figure 13%%%%%%%%%%%%%%%%%%%%%%%%%%%%%%%%%%%%%%
\begin{figure*}
\centering   
	\includegraphics[width=15.2cm]{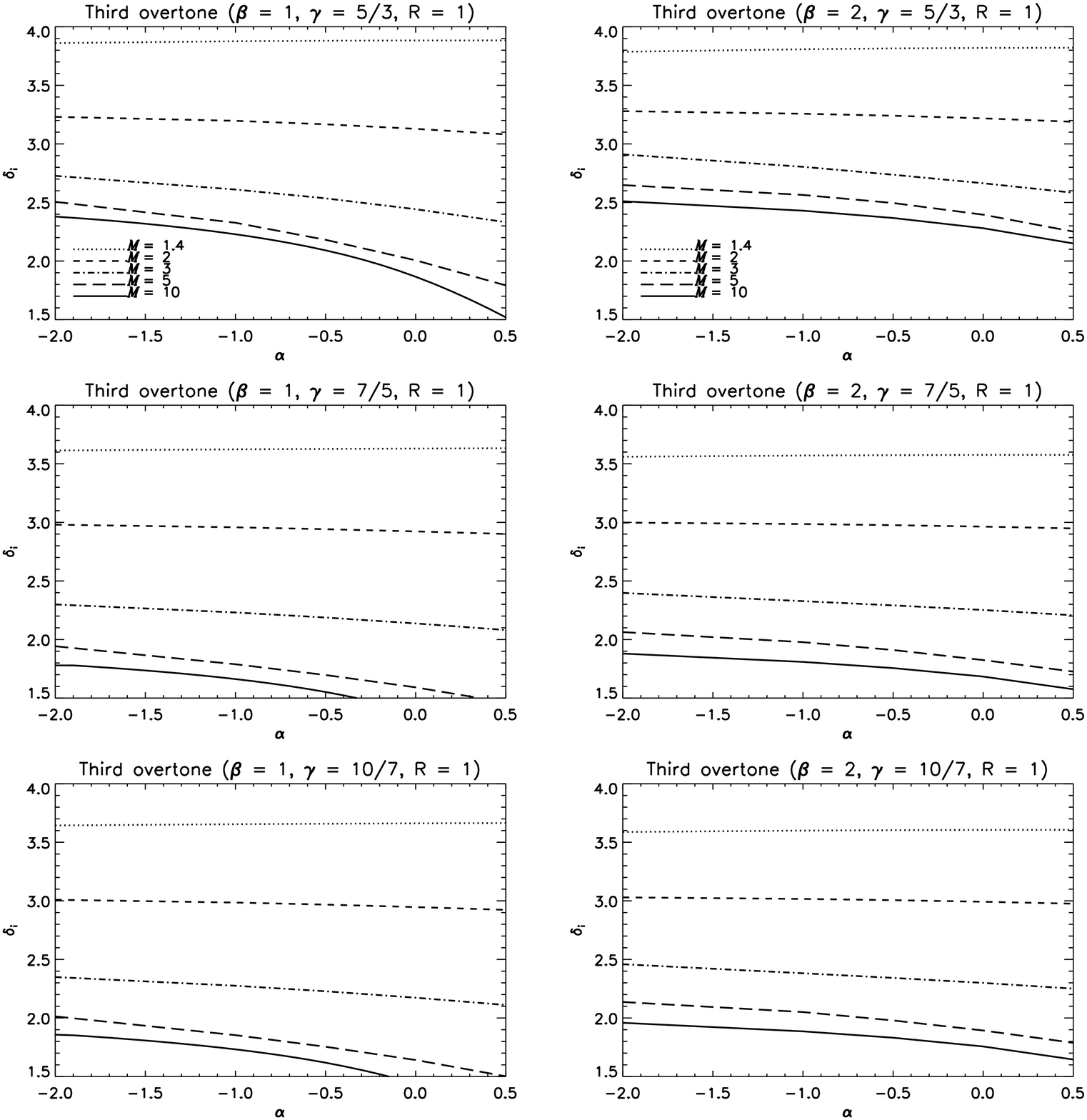} 
\caption{ The eigenfrequency, $\delta_{i}$ as a function of $\alpha$ for the 
third overtone mode  with R\,=\,1. Values of $\gamma$, $\beta$  and $M$ are indicated.}
\label{thirdiR1}
\end{figure*} 
%%%%%%%%%%%%%%%%%%%%%%%%%%%%%%Figure 13%%%%%%%%%%%%%%%%%%%%%%%%%%%%%%%%%%%%%%%

%%%%%%%%%%%%%%%%%%%%%%%%%%%%%%%%%%%%%%%%%%%%%%%%%%%%%%%%%%%%%%%%%%%%%%%%%%%%%%
\section{Method}
%%%%%%%%%%%%%%%%%%%%%%%%%%%%%%%%%%%%%%%%%%%%%%%%%%%%%%%%%%%%%%%%%%%%%%%%%%%%%%
\label{method}

\subsection{The steady state solution}

The method employed here involves searching for wave modes which satisfy the equations representing 
linear perturbations away from the  one-dimensional steady-state solution.
We consider matter of constant density $\rho_{a}$, pressure $P_{a}$, temperature $T_{a}$, sound speed $c_{s}$
and speed $u_{in}$ which is incident on a stationary wall. As sketched in Fig.~\ref{steady}, the 
pre-shock gas velocity is $v = -  u_{in}$ where the shock speed $u_{in}$ is defined as a positive quantity. 
Spatially, the origin  $x = 0$ is located at the wall or shock-shell interface, with the shock front lying at some distance 
$x = x_{s}$. In the case where the final temperature returns to the upstream value, 
the post-shock gas passes through the cooling region and is added to the uniform density shell.
In the case where the final temperature is zero, the matter is added to
an infinitely thin region at the rigid wall. In both cases, the length $x_{s}$ is the thickness of the cooling region.

The one dimensional hydrodynamical equations are
\begin{eqnarray}
  \label{eqmass}
   {\partial \rho \over \partial t} + \rho {\partial v \over \partial x} + 
   v{\partial \rho \over \partial x} &=& 0 , \\
\label{eqmom}
   \rho \left({\partial v \over \partial t} + v{\partial v \over \partial x}\right) 
   + {\partial P \over \partial x} &=& 0,
\end{eqnarray}
\begin{multline}\label{eqenergy}
   {\partial P \over \partial t} + v{\partial P \over \partial x} 
   -  \frac{\gamma P}{\rho} \left({\partial \rho \over \partial t} + 
   v{\partial \rho \over \partial x}\right) = \\
   -(\gamma - 1) A \rho^{\beta - \alpha} P^{\alpha} ,
\end{multline}
where equations (\ref{eqmass}), (\ref{eqmom}) \& (\ref{eqenergy}) refer to the conservation 
of mass, momentum and energy, respectively.
In equation~(\ref{eqenergy}), $A$ is a constant. The steady 
state solution is denoted by the subscript 0. 

The Rankine-Hugoniot conditions provide the values of the physical variables at $x = x_{s}$. These are 
\begin{eqnarray}
\label{rhozero}
   \rho_{0}(x_{s}) &=& \left[\frac {\left(\gamma + 1\right)M^{2}} {\left (\gamma - 1 \right)M^{2} + 2} \right]\rho_{a},\\
\label{velzero}
   v_{0}(x_{s}) &=& - \left[ \frac {\left (\gamma - 1 \right)M^{2} + 2} {\left( \gamma + 1\right)M^{2}} \right]u_{in},\\
\label{pzero}
   P_{0}(x_{s}) &=&  \left[\frac {2\gamma M^{2} - \left(\gamma - 1 \right)} {\gamma \left(\gamma + 1 \right) M^{2}}\right]\rho_{a}u_{in}^{2} .
\end{eqnarray}
\noindent
where $M = \frac{u_{in}}{c_{s}}$ is the Mach number and $c_{s}$ is the sound speed \citep[e.g.][]{Priest,Shore}. 
Equations (\ref{rhozero}), (\ref{velzero}) \& (\ref{pzero}) are obtained when
equations  (\ref{eqmass}) and (\ref{eqmom}) are integrated to yield
\begin{eqnarray}
\label{7}
   \rho_{0}v_{0} &=& - \rho_{a}u_{in},\\
\label{8}
   P_{0} &=& \rho_{a} u_{in}(v_{0} + u_{in}) + P_{a} .
\end{eqnarray}
Equations (\ref{7}) \& (\ref{8}) are substituted in equation (\ref{eqenergy})
 to obtain
\begin{eqnarray}
\label{9}
\frac{dv_{0}}{dx} = \frac{-(\gamma - 1)A(\rho_{a} u_{in})^{\beta - 1} \left[- v_{0}^{2}- u_{in}v_{0} \left ( 1 + \frac {1} { \gamma M^{2} } \right)\right]^{\alpha}} {\left (-v_{0}\right)^{\beta}\left[v_{0} + 
\gamma\left(v_{0} + u_{in} \left ( 1 + \frac {1} {\gamma M^{2}} \right)\right) \right]} .
\end{eqnarray}
We now introduce the following variables 
\begin{eqnarray}
\label{xi}
\xi &=& \frac{x}{x_{s}},\\
\label{w}
w &=& \frac{v_{0}}{u_{in}}.
\end{eqnarray}
Equations (\ref{9}), (\ref{xi}) and (\ref{w}) lead to 
\begin{eqnarray}
\label{10}
\frac{d\xi}{dw} &=& 
\frac {-(-w)^{\beta} u_{in}^{3 - 2\alpha}\left[w + \gamma \left (w + 1 + \frac {1} {\gamma M^{2}} \right)\right]
}{(\gamma - 1)A\rho_{a}^{\beta - 1}x_{s}\left[-w \left (w + 1 + \frac {1} {\gamma M^{2}} \right)\right]^{\alpha}}.
\end{eqnarray}

%%%%%%%%%%%%%%%%%%%%%%%%%%%%%%%%%%%%%%%%%%%%%%%%%%%%%%%%%%%%%%%%%%%%%%%%%%%
\subsection{The steady state boundary conditions}

Before we can integrate equation~(\ref{10}), we must specify the boundary conditions. We consider 
two scenarios. The first case is when the temperature at the shell is the
same as the temperature of the pre-shock medium and the second case is when the 
temperature drops to zero. Therefore, we define a quantity $R$ as the ratio
of the temperature at the shell to that of the pre-shock temperature. If
$R = 1$, then we have the former case and if $R = 0$, we have the latter.

\begin{eqnarray}
R =   \frac {\left (\frac{P_{0}}{\rho_{0}}\right)}{\left ( \frac{P_{a}}{\rho_{a}}\right)}.
\end{eqnarray}
This results in
\begin{eqnarray}
w^{2} + w\left[ 1 + \frac{1}{\gamma M^{2}} \right] + \frac{R}{\gamma M^{2}}=0
\end{eqnarray}
We consider only the roots that are physical and hence the boundary conditions are
\begin{eqnarray}
\label{11}
w &=& 0 \hskip 3.cm {\rm at} \hskip 0.4cm \xi = 0 \hskip 0.4cm (R = 0) \\
\label{12}
w &=& -\left(\frac {1}{\gamma M^{2}}\right)  \hskip 1.6cm {\rm at} \hskip 0.4cm \xi = 0 \hskip 0.4cm (R = 1) \\
\label{13}
w &=& -\left[ \frac {\left (\gamma - 1 \right)M^{2} + 2} {\left (\gamma + 1 \right)M^{2}}\right] \hskip 0.3cm {\rm at} \hskip 0.4cm \xi = 1.
\end{eqnarray}

\noindent
The shock width $x_{s}$ is evaluated in the process.

%%%%%%%%%%%%%%%%%%%%%%%%%%%%%%Figure 13%%%%%%%%%%%%%%%%%%%%%%%%%%%%%%%%%%%%%%%
%%%%%%%%%%%%%%%%%%%%%%%%%%%%%%Figure 14%%%%%%%%%%%%%%%%%%%%%%%%%%%%%%%%%%%%%%%
\begin{figure*}
\centering   
	\includegraphics[width=15.2cm]{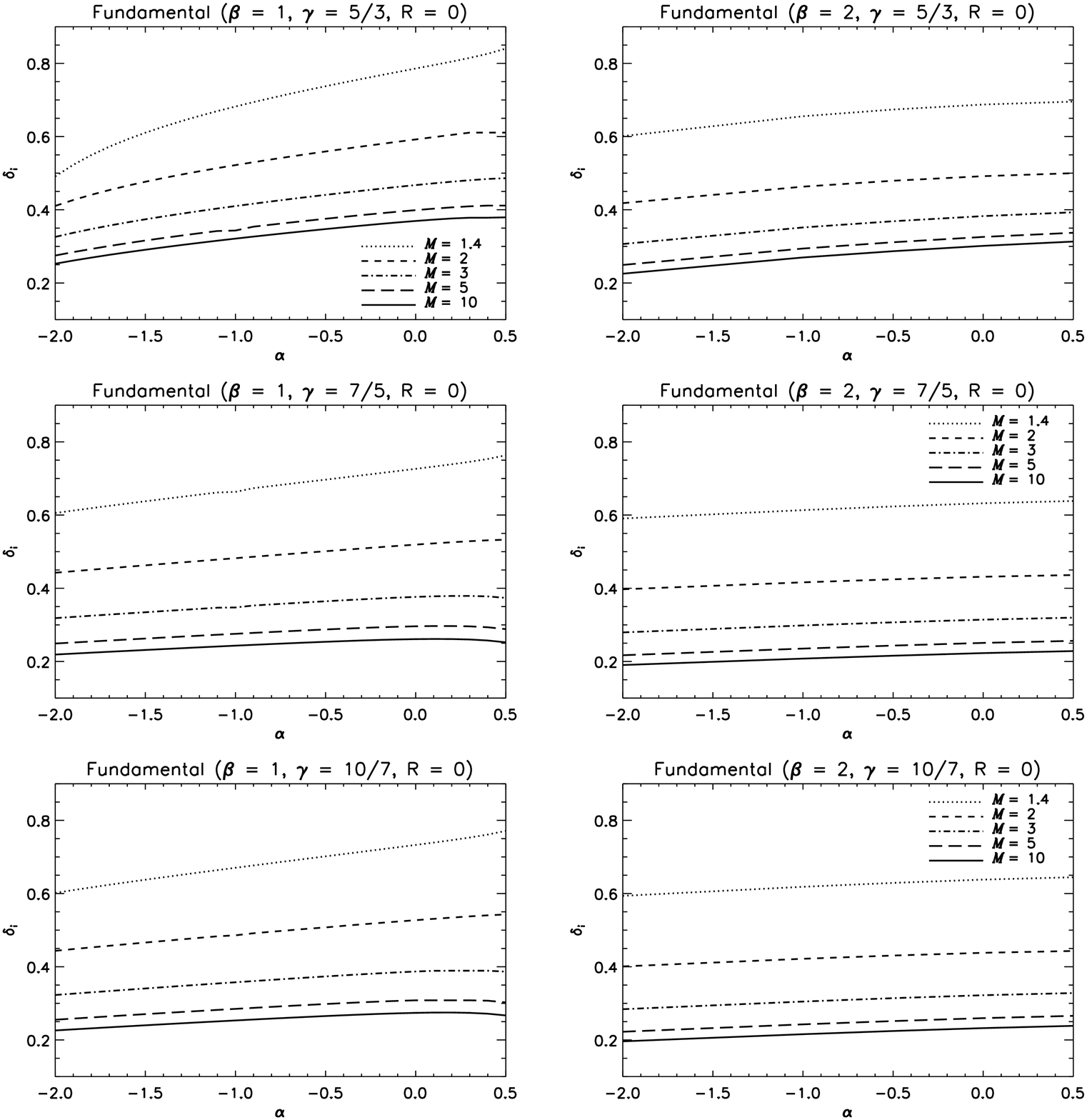} 
\caption{ The eigenfrequency, $\delta_{i}$ as a function of $\alpha$ for the 
fundamental mode with R\,=\,0. Values of $\gamma$, $\beta$  and $M$ are indicated.}
\label{fundiR0}
\end{figure*} 
%%%%%%%%%%%%%%%%%%%%%%%%%%%%%%Figure 14%%%%%%%%%%%%%%%%%%%%%%%%%%%%%%%%%%%%%%%
%%%%%%%%%%%%%%%%%%%%%%%%%%%%%%Figure 15%%%%%%%%%%%%%%%%%%%%%%%%%%%%%%%%%%%%%%%
\begin{figure*}
\centering   
	\includegraphics[width=15.2cm]{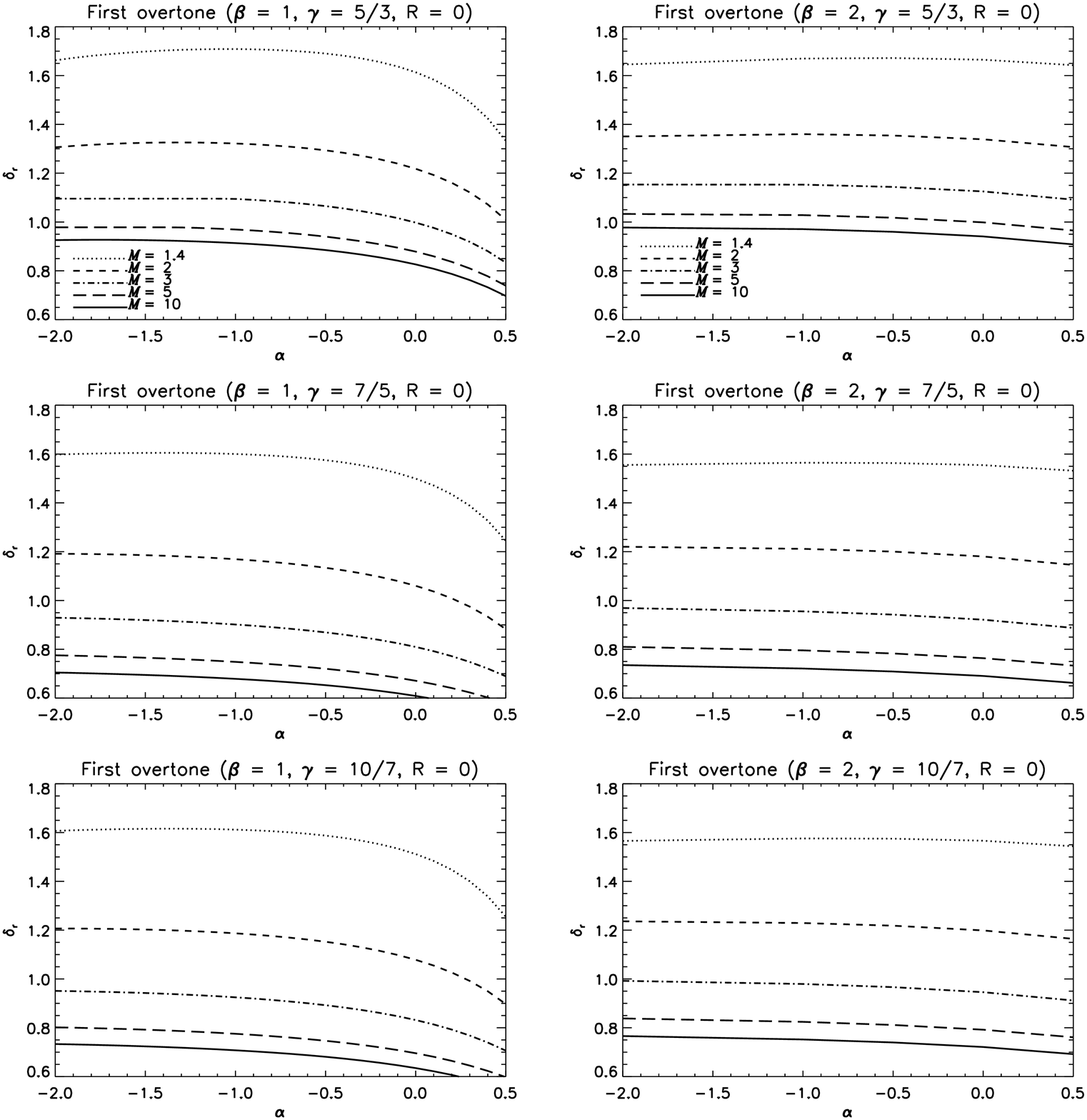} 
\caption{ The eigenfrequency, $\delta_{i}$ as a function of $\alpha$ for the 
first overtone mode with R\,=\,0. Values of $\gamma$, $\beta$  and $M$ are indicated.}
\label{firstiR0}
\end{figure*} 
%%%%%%%%%%%%%%%%%%%%%%%%%%%%%%Figure 15%%%%%%%%%%%%%%%%%%%%%%%%%%%%%%%%%%%%%%%
%%%%%%%%%%%%%%%%%%%%%%%%%%%%%%Figure 16%%%%%%%%%%%%%%%%%%%%%%%%%%%%%%%%%%%%%%%
\begin{figure*}
\centering   
	\includegraphics[width=15.2cm]{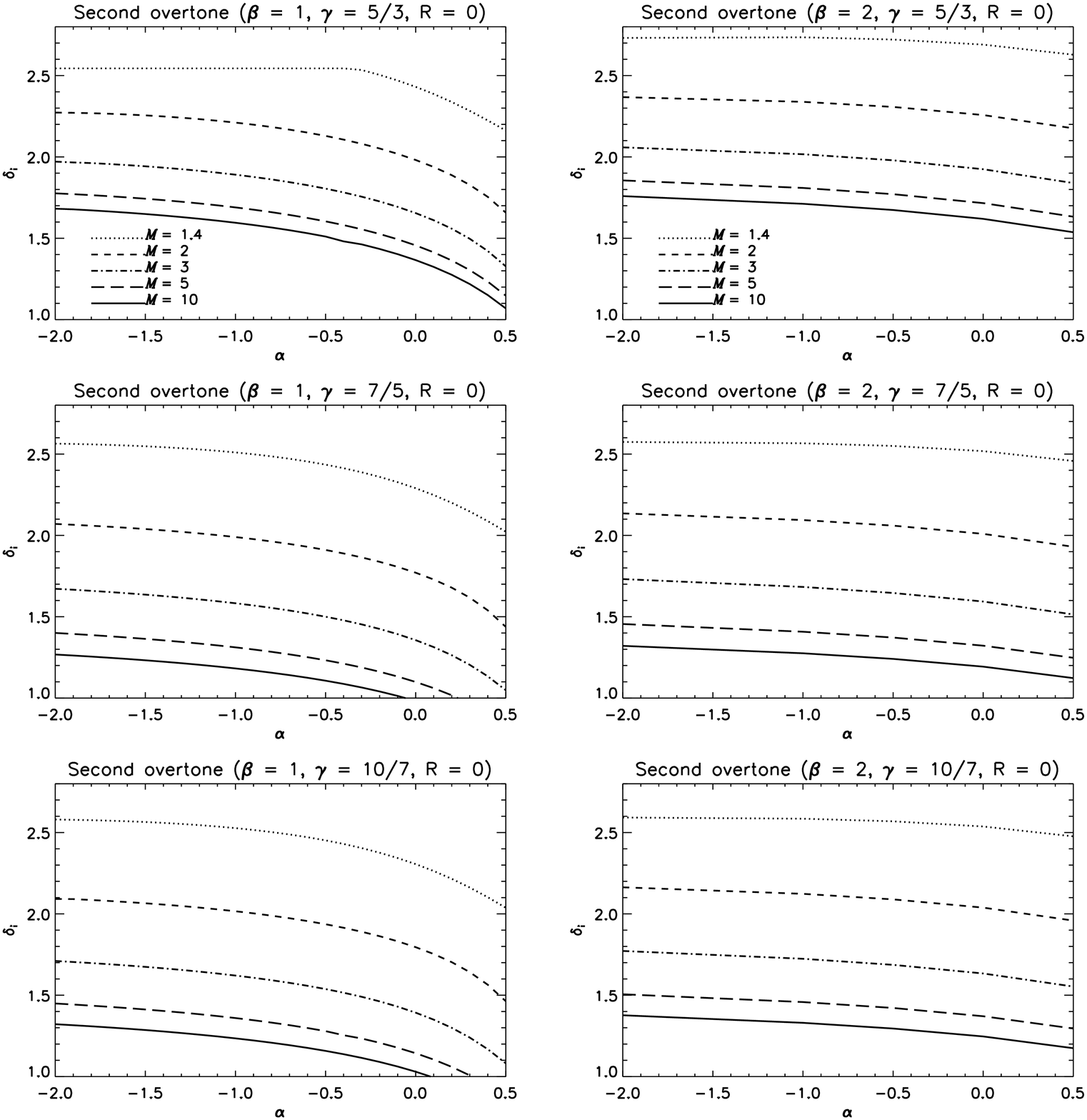} 
\caption{ The eigenfrequency, $\delta_{i}$ as a function of $\alpha$ for the 
second overtone with R\,=\,0. Values of $\gamma$, $\beta$  and $M$ are indicated.}
\label{seciR0}
\end{figure*} 
%%%%%%%%%%%%%%%%%%%%%%%%%%%%%%Figure 16%%%%%%%%%%%%%%%%%%%%%%%%%%%%%%%%%%%%%%%
%%%%%%%%%%%%%%%%%%%%%%%%%%%%%%Figure 17%%%%%%%%%%%%%%%%%%%%%%%%%%%%%%%%%%%%%%%
\begin{figure*}
\centering   
	\includegraphics[width=15.2cm]{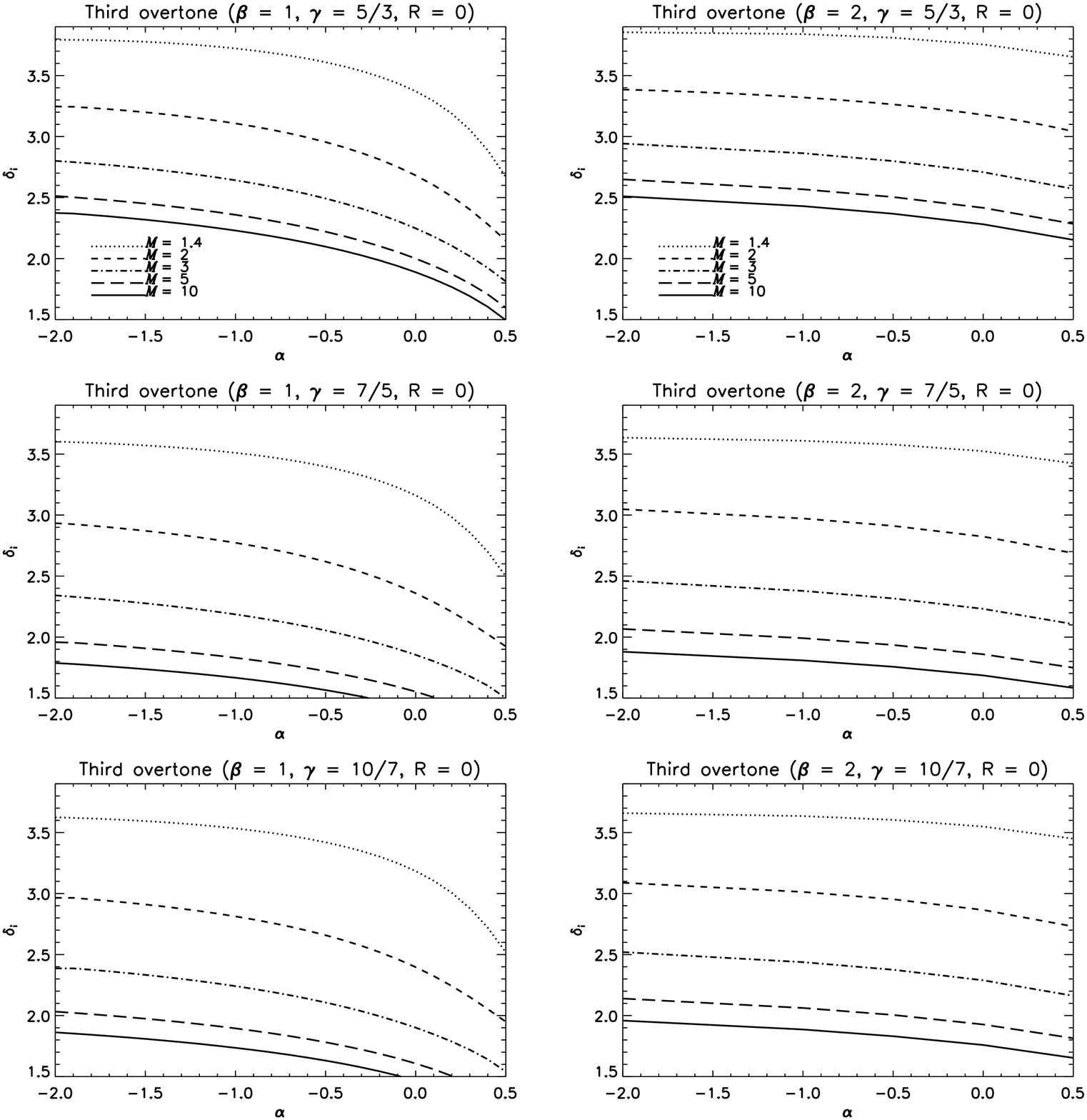} 
\caption{ The eigenfrequency, $\delta_{i}$ as a function of $\alpha$ for the 
third overtone  with R\,=\,0. Values of $\gamma$, $\beta$  and $M$ are indicated.}
\label{thirdiR0}
\end{figure*} 
%%%%%%%%%%%%%%%%%%%%%%%%%%%%%%Figure 17%%%%%%%%%%%%%%%%%%%%%%%%%%%%%%%%%%%%%%%

%%%%%%%%%%%%%%%%%%%%%%%%%%%%%%Figure 18%%%%%%%%%%%%%%%%%%%%%%%%%%%%%%%%%%%%%%%
\begin{figure*}
\centering   
	\includegraphics[width=15.2cm]{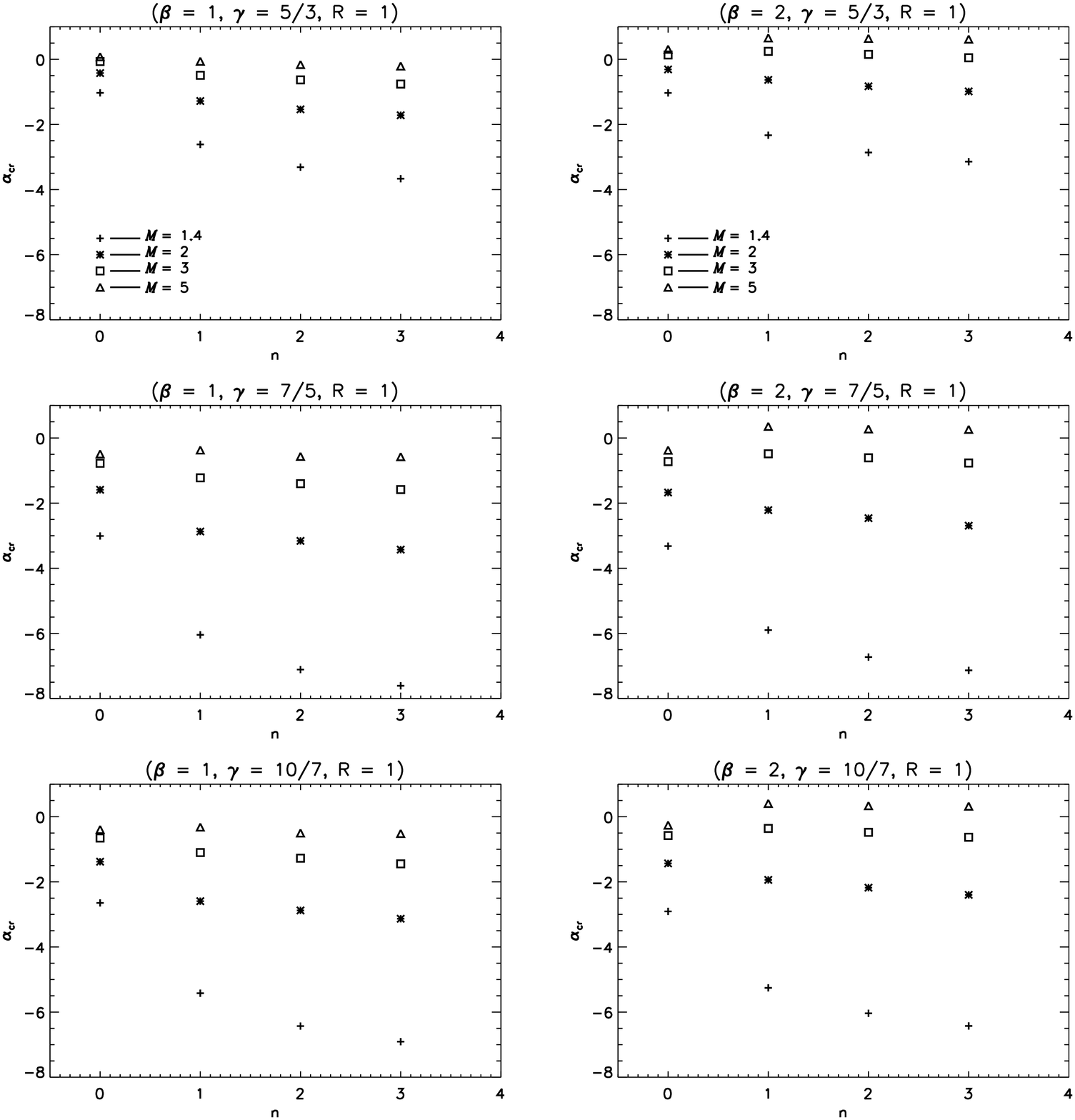} 
\caption{ The critical values of $\alpha$ plotted against the overtones 
 for the displayed values of $\gamma$, $\beta$ and $M$, with  R\,=\,1.}
\label{criticalR1}
\end{figure*} 
%%%%%%%%%%%%%%%%%%%%%%%%%%%%%%Figure 18%%%%%%%%%%%%%%%%%%%%%%%%%%%%%%%%%%%%%%%
%%%%%%%%%%%%%%%%%%%%%%%%%%%%%%Figure 19%%%%%%%%%%%%%%%%%%%%%%%%%%%%%%%%%%%%%%%
\begin{figure*}
\centering   
	\includegraphics[width=15.2cm]{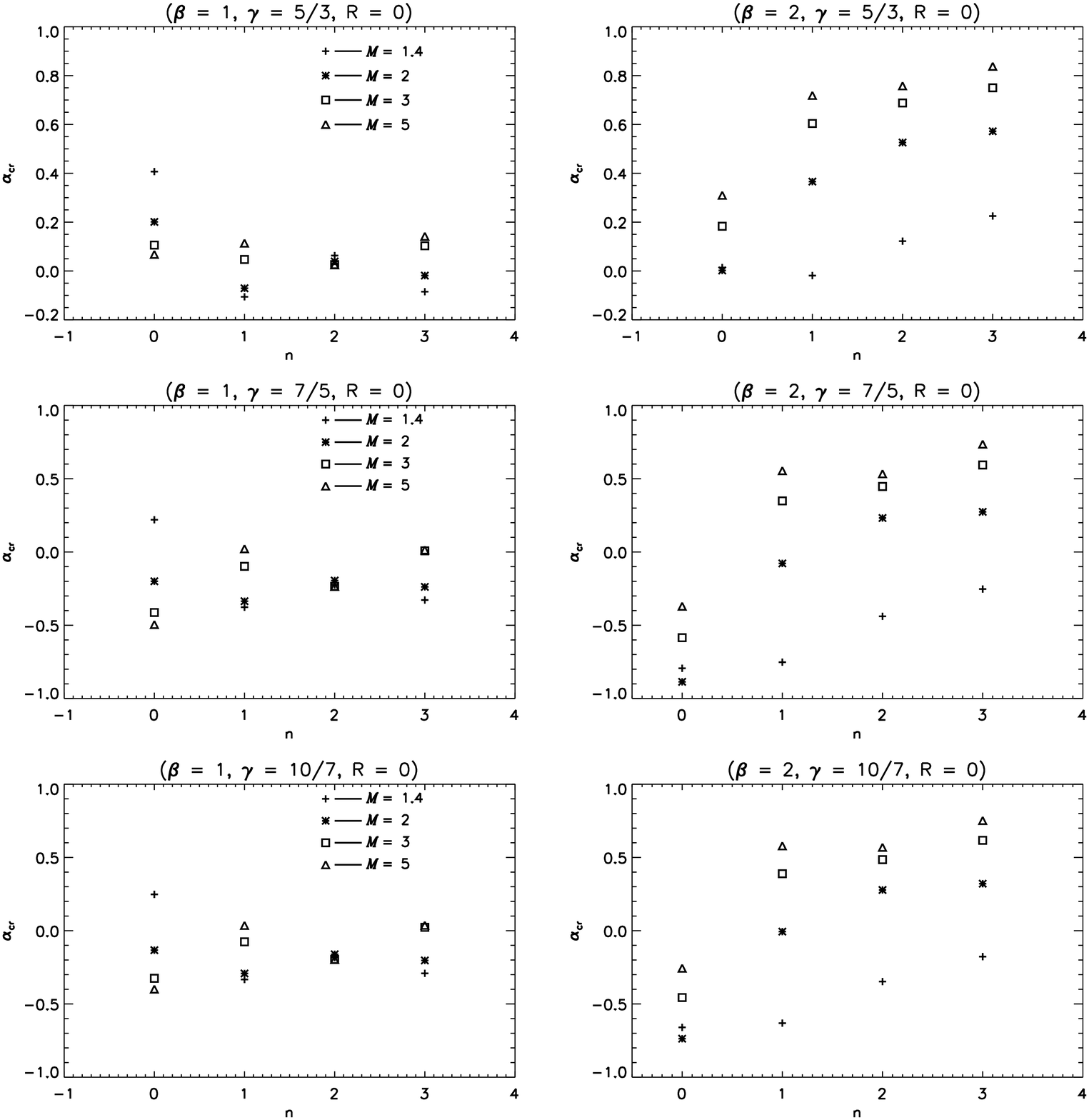} 
\caption{ The critical values of $\alpha$ plotted against the overtones 
 for the displayed values of $\gamma$, $\beta$  and $M$ with R\,=\,0.}
\label{criticalR0}
\end{figure*} 
%%%%%%%%%%%%%%%%%%%%%%%%%%%%%%Figure 19%%%%%%%%%%%%%%%%%%%%%%%%%%%%%%%%%%%%%%%

%%%%%%%%%%%%%%%%%%%%%%%%%%%%%%%%%%%%%%%%%%%%%%%%%%%%%%%%%%%%%%%%%%%
\section{The set of linear equations}
%%%%%%%%%%%%%%%%%%%%%%%%%%%%%%%%%%%%%%%%%%%%%%%%%%%%%%%%%%%%%%%%%%%

The shock wave is now perturbed by
\begin{eqnarray}
\label{13a}
\frac{dx_{s}}{dt} &=& v_{s1}e^{\sigma t},
\end{eqnarray}
\noindent
where $\sigma = \sigma_{R} + i\sigma_{I}$ is the frequency and $v_{s1}$ is a real quantity.
The position of the shock may be represented as the real part of
\begin{eqnarray}
\label{14}
x_{s} &=& x_{s0} + x_{s1}e^{\sigma t},
\end{eqnarray}
where $x_{s1} = \frac{v_{s1}}{\sigma}$. Considering only the terms up to first order:
\begin{eqnarray}
\label{15}
\xi = \frac{x}{x_{s}} &=& \frac{x}{x_{s0}} \left(1 -\frac{x_{s1}}{x_{s0}}e^{\sigma t}
 \right),\\
\label{16}
{\partial \xi \over \partial x} &=&  \frac{1}{x_{s0}}
\left(1 - \frac{x_{s1}}{x_{s0}}e^{\sigma t}\right),  \\
\label{17}
{\partial \xi \over \partial t} &=&  -\frac{x x_{s1} \sigma e^{\sigma t}}{x_{s0}^{2}}, \\
\label{18}
\rho &=& \rho_{0}(\xi) + \rho_{1}(\xi)e^{\sigma t},\\
\label{19}
P &=& P_{0}(\xi) + P_{1}(\xi)e^{\sigma t},\\
\label{20}
v &=& v_{0}(\xi) + v_{1}(\xi)e^{\sigma t}.
\end{eqnarray}
All the quantities with subscript 1 represent the small perturbed factors.
The boundary conditions at the shock wave are (see Appendix~\ref{boundshock})
\begin{eqnarray}
\label{21}
\rho_{1} &=& \left [ \frac{4 v_{s1} \rho_{a} \left(\gamma + 1 \right)M^{2} } {u_{in}\left[ 2 + \left(\gamma - 1 \right) M^{2}\right]^{2}} \right], \\
\label{22}
P_{1}  &=& \left(\frac{4u_{in}v_{s1}\rho_{a}}{\gamma + 1}\right),\\
\label{23}
v_{1}  &=& \frac{2v_{s1}}{\gamma + 1} \left[ 1 + \frac{1}{M^{2}} \right].
\end{eqnarray}

We then transform the following variables as 
\begin{eqnarray}
\label{24}
\zeta &=& \frac{x_{s0}\sigma \rho_{1}}{v_{s1}\rho_{a}}\\
\label{25}
\pi &=& \frac{P_{1}}{v_{s1}\rho_{a}u_{in}}\\
\label{26}
\eta &=& \frac{v_{1}}{v_{s1}}\\
\label{27}
\delta &=& \frac{x_{s0}\sigma}{u_{in}}.
\end{eqnarray}
Substituting (\ref{18}), (\ref{19}), (\ref{20}) into (\ref{eqmass}),  (\ref{eqmom}) and
(\ref{eqenergy}), 
the fluid equations become\\
%\vskip -.40cm
\begin{eqnarray} 
\label{28}
-\frac {\xi}{w^{2}} + \zeta \frac{d\xi}{dw} + \frac{w}{\delta}\frac{d\zeta}{dw} + 
\frac{\eta}{w^{2}} -\frac{1}{w} \frac{d\eta}{dw} + \frac{\zeta}{\delta} &=& 0\\
\label{29}
-\xi + \delta \frac{d\xi}{dw}\eta + w\frac{d\eta}{dw} + \eta - \frac{w^{2}}{\delta}\zeta -
 w\frac{d\pi}{dw} &=& 0\\
%\end{eqnarray}
%\vskip -1.0cm
%\begin{eqnarray}
\label{30}
D + E &=& F
\end{eqnarray} 
where
\begin{eqnarray}
D &=& \left[-\xi  + \delta \frac{d\xi}{dw}\pi + w\frac{d\pi}{dw} +\eta - \frac{w}{\delta} + 
\pi \gamma \right]\nonumber\\
E &=& \left(-\frac {\xi}{w^{2}} + 
\zeta \frac{d\xi}{dw} + \frac{w}{\delta}\frac{d\zeta}{dw} + \frac{\eta}{w^{2}} 
-\frac{1}{\delta w}  + \frac{\zeta}{\delta} \right)\nonumber\\ &&\times \left[\gamma w \left (w + 1 + \frac{1}{\gamma M^{2}}\right)\right] \nonumber\\
F &=&\left[w + \gamma \left ( 1 + w +\frac{1}{\gamma M^{2}}\right)\right]\\ &&\times \left[\frac{\alpha \pi}{\left ( 1 + w + \frac{1}{\gamma M^{2}}\right) } - 
(\beta - \alpha)\frac{w\zeta}{\delta}\right] \nonumber
\end{eqnarray} 
%\vskip -2.5cm
\noindent
The quantities $\zeta$, $\pi$ and $\eta$ are complex eigenfunctions where the subscript $r$ denotes the real 
component and $i$ stands for the imaginary part for each of the above quantities. The quantity $\delta$ is 
 a complex number with the sign of the  real part, $\delta_{r}$ indicating  
the instability (+ve value) or stability (-ve value) of a mode.
The quantity $\delta_{i}$ is interpreted as the eigenfrequency (in units 
of ($u_{in}/x_{s0}$)). 

Substituting  Eqs. (\ref {25}), (\ref {26}) and (\ref {27}) in Eqns. (\ref{28}), (\ref{29}) and (\ref{30}), we get six coupled first order equations which are
\begin{eqnarray} 
\label{31}
\frac{d\eta_{r}}{dw} &=& \frac {\alpha \pi_{r}}{\left ( w + 1 + \frac{1}{\gamma M^{2}} \right)} - \frac {w(\beta - \alpha)(\delta_{r}\zeta_{r} + \delta_{i}\zeta_{i})}{\delta^{2}} + \frac{\delta_{r}}{\delta^{2}}\nonumber\\ && + \frac{2\xi \delta^{2} - 2\eta_{r}  \delta^{2} - \gamma\pi_{r} \delta^{2} + w^{2}(\zeta_{r}\delta_{r} + \zeta_{i}\delta_{i})}{\left[w + \gamma \left(w + 1 + \frac{1}{\gamma M^{2}}\right)\right]\delta^{2}}\nonumber\\ && -\frac{d\xi}{dw}\left[\frac{\delta_{r}(\eta_{r} + \pi_{r}) - \delta_{i}(\eta_{i} + \pi_{i})}{\left(w + \gamma \left (w + 1 + \frac{1}{\gamma M^{2}}\right )\right)}\right]\\
\label{32}
\frac {d\eta_{i}}{dw} &=& \frac {\alpha \pi_{i}}{\left (w + 1 + \frac{1}{\gamma M^{2}} \right)} - \frac {w(\beta - \alpha)(\delta_{r}\zeta_{i} - \delta_{i}\zeta_{r})}{\delta^{2}} - \frac{\delta_{i}}{\delta^{2}}\nonumber\\ && + \frac{- 2\eta_{i}\delta^{2} - \gamma\pi_{i} \delta^{2} + w^{2}(\zeta_{i}\delta_{r} - \zeta_{r}\delta_{i})}{\left[w + \gamma \left (w + 1 + \frac{1}{\gamma M^{2}}\right )\right]\delta^{2}}\nonumber\\ && -\frac{d\xi}{dw}\left[\frac{\delta_{i}(\eta_{r} + \pi_{r}) + \delta_{r}(\eta_{i} + \pi_{i})}{\left(w + \gamma \left (w + 1 + \frac{1}{\gamma M^{2}}\right )\right)}\right]
\end{eqnarray}
\begin{eqnarray}
\label{33}
\frac {d\pi_{r}}{dw} &=& \frac {-\xi}{w} + \frac {(\delta_{r}\eta_{r} - \delta_{i}\eta_{i})}{w} \frac {d\xi}{dw} + \frac {\eta_{r}}{w} + \frac {d\eta_{r}}{dw}\nonumber\\ && - \frac{w(\zeta_{r}\delta_{r} + \zeta_{i}\delta_{i})}{\delta^{2}}
\end{eqnarray}
\begin{eqnarray}
\label{34}
\frac {d\pi_{i}}{dw} &=& \frac {(\delta_{i}\eta_{r} + \delta_{r}\eta_{i})}{w} \frac {d\xi}{dw} + \frac {\eta_{i}}{w} + \frac {d\eta_{i}}{dw}\nonumber\\ && - \frac{w(\zeta_{i}\delta_{r} - \zeta_{r}\delta_{i})}{\delta^{2}}
\end{eqnarray}
\begin{eqnarray}
\label{35}
\frac {d\zeta_{r}}{dw} &=& \frac {\xi\delta_{r}}{w^{3}} - \frac {d\xi}{dw}\frac{\zeta_{r}\delta_{r}}{w}  + \frac {\delta_{r}}{w^{2}} \frac {d\eta_{r}}{dw} - \frac {\eta_{r}\delta_{r}}{w^{3}}\nonumber\\ && - \frac {\delta_{r}(\zeta_{r}\delta_{r} + \zeta_{i}\delta_{i})}{w\delta^{2}} +\frac {d\xi}{dw}\frac{\zeta_{i}\delta_{i}}{w} - \frac {\delta_{i}}{w^{2}} \frac {d\eta_{i}}{dw}\nonumber\\ && + \frac {\eta_{i}\delta_{i}}{w^{3}} + \frac {\delta_{i}(\zeta_{i}\delta_{r} - \zeta_{r}\delta_{i})}{w\delta^{2}}\\
\label{36}
\frac {d\zeta_{i}}{dw} &=& \frac {\xi\delta_{i}}{w^{3}} - \frac {d\xi}{dw}\frac{\zeta_{i}\delta_{r}}{w}  + \frac {\delta_{r}}{w^{2}} \frac {d\eta_{i}}{dw} - \frac {\eta_{i}\delta_{r}}{w^{3}}\nonumber\\ && - \frac {\delta_{r}(\zeta_{i}\delta_{r} - \zeta_{r}\delta_{i})}{w\delta^{2}} - \frac {d\xi}{dw}\frac{\zeta_{r}\delta_{i}}{w} + \frac {\delta_{i}}{w^{2}} \frac {d\eta_{r}}{dw}\nonumber\\ && - \frac {\eta_{r}\delta_{i}}{w^{3}} - \frac {\delta_{i}(\zeta_{r}\delta_{r} + \zeta_{i}\delta_{i})}{w\delta^{2}}
\end{eqnarray} 

\noindent 
There are now four free parameters: $\alpha$, $\beta$, $\gamma$ and $M$. The quantities
 $\delta_{r}$ and $\delta_{i}$ are eigenvalues which are determined by 
imposing the boundary condition at the wall which is
 $|\pi w M\sqrt \gamma - \eta|$ = 0 (see Appendix~\ref{boundshell}).
We solved the differential equations employing a fourth order Runge-Kutta 
technique for trial values of $\delta_{r}$ and $\delta_{i}$ with the 
combination that satisfies the boundary condition being the eigenvalue. We followed the method described by 
\citet{1998MNRAS.299..862S} which involves choosing a grid of points in the complex plane consisting 
of $\delta_{r}$ and $\delta_{i}$ and integrating the equations for each point on the grid. The combinations that come closest to satisfying the boundary condition for each mode determine a new set of grid points with a higher resolution.

%%%%%%%%%%%%%%%%%%%%%%%%%%%%%%%%%%%%%%%%%%%%%%%%%%%%%%%%%%%%%%%%%%%%%%%%%%%%%%%%%%%%%%%%%%%%%%%%%%%%%%%%%%%%%%%%%%%%
\section{Results}
%%%%%%%%%%%%%%%%%%%%%%%%%%%%%%%%%%%%%%%%%%%%%%%%%%%%%%%%%%%%%%%%%%%%%%%%%%%%%%%%%%%%%%%%%%%%%%%%%%%%%%%%%%%%%%%%%%%%
\label{results}

The main result is that a finite Mach number alters the
regime of overstability as compared to the strong shock limit.
However, the alteration can be in either direction.

For the standard hot atomic case, the weaker the shock,
the lower the critical temperature power-law index $\alpha_c$ below which shocks are overstable
to the fundamental mode (with $\beta = 2$ and $\gamma = 5/3$ fixed). This case is displayed in the top-right
panels of Figures \ref{fundrR1} \& \ref{fundrR0} for the two extremes $R = 1$ and  $R = 0$, respectively.
This is also true for the overtones (see top-right panels of Figs.~(\ref{firstrR1}\,--\,\ref{thirdrR1}) and 
Figs.~(\ref{firstrR0}\,--\,\ref{thirdrR0}). Thus, the figures
demonstrate that the major long wavelength modes are significantly more stable.  

For the fundamental mode, the stabilising effect is of moderate significance. 
Even for $M = 3$ the critical index $\alpha_c$
is typically reduced by only 0.1\,--\,0.2 for all cases with $R = 1$. For the overtones, however,
the stabilising effect is greater with  $\alpha_c$ reduced by $\sim 0.4$.

Similar results hold for the diatomic/molecular case ($\gamma=7/5$) and the molecular+helium case
($\gamma=10/7$), as illustrated in the corresponding (lower-right and middle-right) panels, 
for $\beta = 2$. The figures illustrate once again that there is no significant difference if the degrees of freedom of
the helium atoms are taken into account.

Remarkably, for the  $\beta = 1$ case, the fundamental mode can grow within a wider range of  $\alpha_c$
values. That is, the shocks become more unstable due to the finite Mach number (see all the left
panels of Figs. \ref{fundrR1} \& \ref{fundrR0}). This result does not hold for
the overtones. The increased instability range is systematically and strongly present for the case $R = 0$ and $\beta = 1$
(left panels of Fig. \ref{fundrR0}). For  $R = 1$, however, a slightly wider range exists only for moderate Mach numbers
of order $\sim 5$.

In fact, even for  $\beta = 2$, although the instability regime is not
widened, growth rates are increased relative to the strong shock for sufficiently low/negative
temperature indices $\alpha$ (illustrated by the crossing of the loci for the growth rates in the
figures). This result holds for the overtones also.

These results can be interpreted in terms of two competing physical effects. Firstly, as with the 
damping magnetic field, the finite Mach number cushions the shock by allowing faster sound wave propagation
which should tend to smear out pressure fluctuations.
In addition, for $R = 1$, the transmission of waves at the shell, rather than just reflection at the wall,
provides a stabilising influence.

On the other hand, a reduced Mach number implies less compression and heating at the shock front.
Hence, the front itself behaves in a softer fashion, analogous to a {\em higher} specific heat ratio. This
tends to widen the instability range of   $\alpha_c$, as found in Paper 1. The resistance to the overstability
as $\gamma$ decreases can be interpreted as if the cooling function consisted of two components, with one component with very high $\alpha$
located just in the hot post-shock gas. This extra cooling component (following the shock heating) 
produces the extra compression.

In the case of $R =1$, except for the fundamental mode, the eigenfrequencies tend to have nearly the 
same values. The fundamental mode tends to converge for positive $\alpha$ values for $\beta  = 1$. For $R = 0$ 
scenario, the fundamental shows a mild increase in the eigenfrequencies as $\alpha$ increases.  The overtones have 
almost the same frequency for $\beta = 2$ though for $\beta = 1$ the frequency tends to decrease with increasing  $\alpha$.

In almost all the cases for $\beta$ \& $\gamma$, $R = 0$ scenarios have higher frequencies than $R = 1 $ as well as
the  eigenfrequencies seem to decrease with increasing Mach numbers.

The dependence of the overstable regimes on the mode is further explored in Figs. \ref{criticalR1} \& \ref{criticalR0}.
These figures clearly show that the higher overtones are more sensitive to the
Mach number.
In fact, while the overtones can be exclusively present for a range in $\alpha$ for strong shocks,
the fundamental might be exclusively present for weak shocks. As an example,
we can see in the case of $M = 1.4$ ($\beta = 2$, $\gamma = 5/3$), $\alpha_{c} = -1.031$ for the fundamental mode
whereas the higher overtones have $\alpha_{c} < -2$ (see also Table \ref{b12R1}). Also this significant
change in shock behaviour occurs for Mach numbers below $\sim$ 3 for the case R\,=\,1  (Fig. \ref{criticalR1})
but is more complex for  R\,=\,0 (Fig. \ref{criticalR0}). 

%%%%%%%%%%%%%%%%%%%%%%%%%%%%%%%%%%%%%%%%%%%%%%%%%%%%%%%%%%%%%%%%%%%%%%%%%%%%%%%%%%%%%%%%%%
\section{Conclusions}
%%%%%%%%%%%%%%%%%%%%%%%%%%%%%%%%%%%%%%%%%%%%%%%%%%%%%%%%%%%%%%%%%%%%%%%%%%%%%%%%%%%%%%%%%%

We have derived the linearised equations for one-dimensional weak shocks with general power-law 
cooling functions and specific heat ratios. We then solved the equations numerically and presented tables and 
figures which correspond to a range of conditions relevant to interstellar atomic and molecular
cooling processes. The two cases solved numerically are (1) cooling and accretion onto a cold stationary wall of infinite density and
(2) cooling down to the pre-shock temperature. These two cases should cover most circumstances of interest for
strongly radiative shocks. 

We conclude that for Mach number $M > 5$, the strong shock overstability limits are not significantly modified.
For  $M < 3$, however, shocks are considerably more stable for most cases with the clear exception 
of $\beta = 1$ and $R = 0$ for which greater instability (of the fundamental mode) is found. The stability
criterion for the overtones are more sensitive to the Mach number, which may result in a change in behaviour
from overtone-dominated to fundamental-dominated as the Mach number falls below 3.
These results are roughly consistent with what might be expected since the equations are altered by factors of order
$1/M^2$ in comparison to the strong shock case. We provide a possible explanation for the results in terms 
of a stabilising influence provided downstream but a destabilising effect associated with the shock front. 

We reach the same general conclusion derived from numerical simulations by \citet{2005A&A...438...11P} 
that low Mach number atomic shocks are more stable than in the strong shock limit. However, we do find different results as far as values of 
$\alpha_c$ are concerned, with the numerical work uncovering a wider range of $\alpha$ for overstability. 
As noted by \citet{2005A&A...438...11P}, this may be due to contributions from non-linear effects in the 
numerical simulations.
Their result is interesting especially given that the strong shock limit is closely approximated even for a 
Mach number of 10 (for example, the ratio of the density immediately behind
the shock to the pre-shock density for a strong shock of $\gamma = 5/3$ is 4 in table 1 of \citet{2005A&A...438...11P}). 
Our instability results reveal that the stability criterion is very close to the standard $\alpha_c \sim 0.4 $ 
even for a Mach number of 10 (see  Table \ref{b12R1}). We thus find our linear instability 
results not to be surprising in the sense that the strong shock results are attained very much as expected from the 
Rankine-Hugoniot conditions. 

In terms of shock speeds, our results imply that the stability regime for shock propagation into
an interstellar atomic medium of temperature $\sim$ 10,000\,K must take into account the
finite Mach number only for $v < 50$~km~s$^{-1}$. However, such shocks are in any case expected to be stable 
due to the steep 
temperature dependence of the cooling function. Only if the pre-shock temperature is several times higher than
10,000\,K, can we expect behaviour modified from the strong shock limit, as discussed in detail by \citet{2005A&A...438...11P}.

A further regime of interest is that of the propagation of radiative shocks into cool atomic media. 
Shock waves of speed 10\,--\,15~km~s$^{-1}$ heat atomic gas to temperatures 2,900\,K\,--6,500\,K. Then, 
fine-structure cooling of elements such as C$^+$ and Fe$^+$ dominates the cooling with cooling functions that may
increase as the temperature falls even in scenarios involving non-equilibrium ionisation
\citep{1972ARA&A..10..375D}. Our analysis indicates that such shocks are overstable to the fundamental 
mode even at Mach numbers as low as 3 e.g. even if the temperature in the pre-shock medium is over 1000\,K.
These shocks possess thick radiative layers with relatively long cooling times (of order 10$^5$/n\,yr) 
and correspondingly long growth and non-linear oscillation periods.

For molecular shocks into cold gas, only extremely low 
speed shocks will possess a low Mach number (e.g. $v < 1$~km~s$^{-1}$). Furthermore, we expect the magnetic field to provide
a greater influence on the stability criterion as reported in Paper II. For example, an Alfv\'en (Mach) number of less than 20
will stabilise the fundamental and first three harmonics for all $\alpha$ greater than zero (taking $\beta = 2$).
To achieve the same stability without the magnetic field, requires a Mach number of $M < 4$ for $R = 1$
and  $M <~1.7$ for $R = 0$. Note, however, that we have assumed a transverse magnetic field in Paper II.
 
The combined results of this series of papers indicate that slow molecular jump shocks are generally not overstable 
to the longest-wavelength fundamental mode since
molecular cooling functions tend to be steep positive functions of temperature (see Paper I). The overtones, however,
may be responsible for variability which we suggest will take the form of low-amplitude jittering. 
However, it should be remarked that other hydrodynamic processes involving the dense shell \citep{1994ApJ...428..186V} 
or turbulence in the pre-shock medium \citep{2003ApJ...591..238S} may also drive linear or non-linear instabilities.

Dissociative shocks possess complex time-dependent 
cooling and chemistry within the radiative layer and numerical simulations are necessary to determine the stability 
properties \citep{2003MNRAS.339..133S,2004A&A...427..147L}. In any case,
the possibility to observe the implied quasi-periodic variations depends on the resolution of the observing instrument
and the length of coherence across a shock front to the oscillations. However, even with low resolution,
the development of density structure will still produce shock signatures inconsistent with steady shock models
\citep{2003ApJ...591..238S}. A similar conclusion was drawn from multi-dimensional simulations of
C-type molecular shocks in which ion-neutral streaming generates an instability \citep{1997ApJ...491..596M}.

%%%%%%%%%%%%%%%%%%%%%%%%%%%%%%%%%%%%%%%%%%%%%%%%%%%%%%%%%%%%%%%%%%%%%%%%%%%%%%%%%%%%%%%%%%
\section*{Acknowledgments}
%%%%%%%%%%%%%%%%%%%%%%%%%%%%%%%%%%%%%%%%%%%%%%%%%%%%%%%%%%%%%%%%%%%%%%%%%%%%%%%%%%%%%%%%%%

Research at Armagh Observatory is funded by
the Department of Culture, Arts and Leisure, Northern
Ireland. BR is extremely grateful to Sathya Sai Baba for 
encouragement and also thanks Professor S. Kandaswami
for guidance.

%%%%%%%%%%%%%%%%%%%%%%%%%%%%%%%%%%%%%%%%%%%%%%%%%%%%%%%%%%%%%%%%%%%%%%%%%%%%%%%%%%%
%%%\bibliography{shock}

%%%%%%%%%%%%%%%%%%%%%%%%%%%%%%%%%%%%%%%%%%%%%%%%%%%%%%%%%%%%%%%%%%%%%%%%%%%%%%%%%%%%%
\appendix
%%%%%%%%%%%%%%%%%%%%%%%%%%%%%%%%%%%%%%%%%%%%%%%%%%%%%%%%%%%%%%%%%%%%%%%%%%%%%%%%%%%%%
\section{Boundary conditions at the moving shock}
\label{boundshock}

\noindent
In the steady state case, both the incoming gas and the outgoing gas (at the shock) are the same
for a stationary observer as well as the shock. But the situation changes when the shock starts
oscillating. Now the stationary observer will find the same incoming velocity with a different velocity
for the post-shock gas which consists of two parts, viz, the steady state velocity and a small 
perturbed term which will be determined as follows. Let the velocity of the shock  
be $v_{s}$ = $v_{s1}e^{\sigma t}$ according to the stationary observer. In the frame of the shock,  
\begin{eqnarray}
v_{in} &=&  - u_{in}- v_{s},  
\end{eqnarray}
\noindent
where $v_{in}$  is the incoming gas velocity.  There will be a small perturbation to the Mach number
and the new Mach number ($M_{n}$) is written as  $M_{n} = M + M_{1}$, where 
$M_{1}$ = $\frac {v_{s}}{c_{s}}$ and $M_{1} << M$. Here $c_{s}$ is the velocity of sound. If $v$ is the 
velocity of the post-shock gas in the observer frame, it implies that $v - v_{s}$ will be the velocity as seen from the shock frame. Applying 
the Rankine-Hugoniot condition, we obtain
\begin{eqnarray}
\label{vtotal}
v - v_{s}  &=& \left[\frac {\left ( \gamma - 1 \right) M_{n}^{2} + 2} {\left ( \gamma + 1 \right) M_{n}^{2} }\right]v_{in}
\end{eqnarray}
Rearranging the terms, we obtain
\begin{eqnarray}
v  &=& -\left[\frac {\left ( \gamma - 1 \right) M^{2} + 2} {\left ( \gamma + 1 \right) M^{2} }\right] u_{in} + \frac{2v_{s}}{\gamma + 1}\left[ 1 + \frac{1}{M^{2}} \right].
\end{eqnarray}

\noindent 
Comparing this with equations (\ref{velzero}) and (\ref{20}) we find that the perturbed velocity is
\begin{eqnarray}
v_{1}  &=& \frac{2v_{s1}}{\gamma + 1} \left[ 1 + \frac{1}{M^{2}} \right].
\end{eqnarray}

\noindent 
Now for the perturbation in density, we use the Rankine-Hugoniot condition
which expresses the conservation of mass, viz,
\begin{eqnarray}
\rho_{in}v_{in} &=& \rho \left(v - v_{s} \right),
\end{eqnarray}
where $\rho_{in}$ and $\rho$ are the pre-shock and post-shock densities and
$\rho_{in}$ = $\rho_{a}$. Simplifying, we get 
\begin{eqnarray}
\rho &=& \left[\frac {\left(\gamma + 1\right)M^{2}} {\left (\gamma - 1 \right)M^{2} + 2} \right]\rho_{a}  + \left [ \frac{4 v_{s} \rho_{a} \left(\gamma + 1 \right)M^{2} } {u_{in}\left[ 2 + \left(\gamma - 1 \right) M^{2}\right]^{2}} \right]
\end{eqnarray}
which implies from equations (\ref{rhozero}) and (\ref{18}) that the density perturbation is   
\begin{eqnarray}
\rho_{1} = \left [ \frac{4 v_{s1} \rho_{a} \left(\gamma + 1 \right)M^{2} } {u_{in}\left[ 2 + \left(\gamma - 1 \right) M^{2}\right]^{2}} \right].
\end{eqnarray}

\noindent 
For the pressure perturbation, we use the Rankine-Hugoniot condition for the conservation 
of momentum flux. Denoting $P$ as the post-shocked gas pressure,
\begin{eqnarray}
P &=& \rho_{in}v_{in}^{2} - \rho \left( v - v_{s}\right)^{2} + P_{in},
\end{eqnarray}
where $P_{in} = P_{a}$ is the pre-shock thermal pressure. Solving the above equation results in
\begin{eqnarray}
P &=&  P_{0} + \frac{4\rho_{in}u_{in}v_{s}}{\gamma + 1}, 
\end{eqnarray}
which implies from equation (\ref{19}) that the perturbed part of the pressure is 
\begin{eqnarray}
P_{1} &=& \left(\frac{4\rho_{a}u_{in}v_{s1}}{\gamma + 1}\right).
\end{eqnarray}

%%%%%%%%%%%%%%%%%%%%%%%%%%%%%%%%%%%%%%%%%%%%%%%%%%%%%%%%%%%%%%%%%%%%%%%%%%%%%%%%%%%%%%%%%%%%%%%%%%%%%%%%%%%%%%%%%%%%
\section{Boundary condition at the shell during the shock oscillations}
%%%%%%%%%%%%%%%%%%%%%%%%%%%%%%%%%%%%%%%%%%%%%%%%%%%%%%%%%%%%%%%%%%%%%%%%%%%%%%%%%%%%%%%%%%%%%%%%%%%%%%%%%%%%%%%%%%%%
\label{boundshell}

At the shell we consider the two scenarios. One case is when
the temperature goes to zero. The second is when the temperature
of the fluid cools down  to  the pre-shock temperature.

The first scenario is very simple as the pressure goes to zero, the
density becomes infinite and therefore the velocity goes to zero which implies
the velocity perturbation also vanishes at $w = 0$. Therefore, 
\begin{eqnarray}
\eta = 0 
\end{eqnarray}

The boundary condition for the second case is evaluated as follows.
We represent the physical quantities as  
\begin{eqnarray}
\rho &=& \rho_{0} + \rho_{1}e^{i(ft + kx)},\\
\label{eqp}
P &=& P_{0} + P_{1}e^{i(ft + kx)},\\
v &=& v_{0} + v_{1}e^{i(ft + kx)},
\end{eqnarray}
where $k$ is the wavenumber, which represents the direction of the outgoing wave
towards the shell and $f$ is the frequency. The above equations  are substituted in (\ref{eqmass}) 
and (\ref{eqmom}) to yield
\begin{eqnarray}
\frac{f}{k} = - \left(\frac{\rho_{0} v_{1} v_{0} + P_{1}}{\rho_{0}v_{1}}\right),\\
\frac{f}{k} = - \left(\frac{\rho_{0} v_{1} + \rho_{1}v_{0}}{\rho_{1}}\right).
\end{eqnarray}
Equating the above expressions leads to 
\begin{eqnarray}
\label{p11}
P_{1} = \frac{\rho_{0}^{2} v_{1}^{2}}{\rho_{1}}.
\end{eqnarray}
%The thermal  pressure is given by 
%\begin{eqnarray}
%\label{eqmpp}
%P = \frac{c_{s}^{2}\rho}{\gamma}
%\end{eqnarray}
%where $c_{s}$ is the velocity of sound.
Now the pressure perturbation can also be expressed in terms of the density perturbation by
\begin{eqnarray}
\label{p12}
P_{1} =\frac{c_{s}^{2}\rho_{1}}{\gamma}. 
\end{eqnarray}
Equations (\ref{p11}) and (\ref{p12}) yield the boundary condition at the shell as
\begin{eqnarray}
\label{eqp1}
   P_{1} = - \frac {c_{s}v_{1}\rho_{0}}{\sqrt \gamma}.
\end{eqnarray}
The minus sign indicates the direction of the gas flow. Note that the perturbed quantities are 
functions of $x$, which itself is a function of $\xi$ from (\ref{15}). The expression  
(\ref{eqp1}) can be written in terms of the non-dimensional quantities as
\begin{eqnarray}
\label{eqdp}
\pi = \frac {\eta} { w_{} M \sqrt \gamma}. 
\end{eqnarray}
The above quantities are evaluated at the shell boundary $w = -\frac {1}{\gamma M^{2}}$.

%%%%%%%%%%%%%%%%%%%%%%%%%%%%%%%%%%%%%%%%%%%%%%%%%%%%%%%%%%%%%%%%%%%%%%%%%%%%%%%%%%%%%%%%%%%
\section{Tables} %%%%%%%%%%%%%%%%%%%%%%%%%%%%%%%%%%%%%%%%%%%%%%%%%%%%%%%%%%%%%%%%%%%%%%%%%%
%%%%%%%%%%%%%%%%%%%%%%%%%%%%%%%%%%%%%%%%%%%%%%%%%%%%%%%%%%%%%%%%%%%%%%%%%%%%%%%%%%%%%%%%%%%

%*************************************************************************************************%
\begin{table*} 
 \caption{Growth rates ($\delta_{r}$) for several  modes and   Mach number $M$, for $\gamma = 5/3$, $\beta = 1$, R = 1 }
 \label{g53b1r} %the labelling is gammabeta r
$$
\begin{array}{lccccccccc} 
\hline
\mathsf{ \,\,\,\,\,\,\,\,\,\,\,\,}
\hskip 4.cm \rm fundamental \hskip 6.5cm \rm first \hskip 0.1cm  overtone  \\
\hline
\noalign{\smallskip} \noalign{\smallskip}
	 \mathsf{\alpha \,\,\,\,\,\,\,}
	 \hskip 0.5cm \mathsf{M = 1.4}
	 \hskip 0.55cm \mathsf{M = 2} 
	 \hskip 0.6cm \mathsf{M = 3}
         \hskip 0.7cm \mathsf{M = 5} 
	 \hskip 0.6cm \mathsf{M = 10}
         \hskip 0.6cm \mathsf{M = 40}
	 \hskip 0.7cm \mathsf{M = 1.4}
         \hskip 0.6cm \mathsf{M = 2} 
	 \hskip 0.65cm \mathsf{M = 3}
         \hskip 0.65cm \mathsf{M = 5} 
	 \hskip 0.6cm \mathsf{M = 10}
         \hskip 0.45cm \mathsf{M = 40}
    \\
 \hline  
 \noalign{\smallskip}
\mathsf{-2} \hskip 0.68cm 0.15868\hskip 0.5cm 0.25826  \hskip 0.5cm 0.28033  \hskip 0.5cm  0.24792   \hskip 0.5cm  0.22699 \hskip 0.5cm  0.22036  \hskip 0.4cm  -0.11795  \hskip 0.55cm 0.15932  \hskip 0.57cm 0.35105  \hskip 0.56cm 0.37958  \hskip 0.42cm  0.36678 \hskip 0.4cm  0.36174 \\
 \mathsf{-1}\hskip 0.3cm -0.00505 \hskip 0.52cm 0.09883 \hskip 0.5cm 0.14289  \hskip 0.5cm 0.13754     \hskip 0.5cm 0.12761  \hskip 0.5cm  0.12418   \hskip 0.4cm  -0.32850   \hskip 0.2cm -0.06753  \hskip 0.57cm 0.13312  \hskip 0.56cm 0.21459  \hskip 0.4cm 0.21670  \hskip 0.4cm 0.21518   \\ 
 \mathsf{-0.5} \hskip 0.1cm -0.09487  \hskip 0.5cm 0.01371   \hskip 0.47cm 0.07180 \hskip 0.5cm 0.08018 \hskip 0.52cm 0.07432  \hskip 0.5cm 0.07209   \hskip 0.4cm -0.44397   \hskip 0.2cm -0.19630  \hskip 0.57cm 0.00248   \hskip 0.55cm  0.11434 \hskip 0.4cm 0.13225  \hskip 0.42cm  0.13237  \\ 
 \mathsf{0.0}\hskip 0.3cm -0.19324 \hskip 0.2cm -0.08116 \hskip 0.1cm -0.01084  \hskip 0.5cm 0.01266    \hskip 0.5cm 0.00988   \hskip 0.5cm 0.00782   \hskip 0.38cm  -0.56824   \hskip 0.2cm  -0.33770  \hskip 0.2cm -0.14700 \hskip 0.2cm  -0.01762 \hskip 0.4cm 0.03008  \hskip 0.42cm  0.03384 \\
\end{array}
 $$
\end{table*}
%******************************************************************************%
%%%%%%%%%%%%%%%%%%%%%%%EIGENFREQUENCIES TABLE%%%%%%%%%%%%%%%%%%%%%%%%%%%%%%%%%%
\begin{table*} 
 \caption{Eigenfrequencies ($\delta_{i}$) for several modes and Mach number $M$, for $\gamma = 5/3$, $\beta = 1$, R = 1.}
 \label{g53b1f} %the labelling is gammabeta i
$$
\begin{array}{lccccccccc} 
\hline
\mathsf{ \,\,\,\,\,\,\,\,\,\,\,\,}
\hskip 3.5cm \rm fundamental \hskip 6.5cm \rm first \hskip 0.1cm  overtone  \\
\hline
\noalign{\smallskip} \noalign{\smallskip}
	 \mathsf{\alpha \,\,\,\,\,\,\,}
	 \hskip 0.3cm \mathsf{M = 1.4}
	 \hskip 0.5cm \mathsf{M = 2} 
	 \hskip 0.5cm \mathsf{M = 3}
         \hskip 0.5cm \mathsf{M = 5} 
	 \hskip 0.5cm \mathsf{M = 10}
         \hskip 0.5cm \mathsf{M = 40}
	 \hskip 0.7cm \mathsf{M = 1.4}
         \hskip 0.65cm \mathsf{M = 2} 
	 \hskip 0.65cm \mathsf{M = 3}
         \hskip 0.5cm \mathsf{M = 5} 
	 \hskip 0.7cm \mathsf{M = 10}
        \hskip 0.5cm \mathsf{M = 40}

    \\
 \hline  
 \noalign{\smallskip}
\mathsf{-2}\hskip 0.5cm 0.49769\hskip 0.5cm 0.38303  \hskip 0.4cm 0.31619  \hskip 0.3cm  0.27455   \hskip 0.3cm  0.25269 \hskip 0.5cm  0.24564 \hskip 0.7cm  1.56673 \hskip 0.5cm 1.25451  \hskip 0.5cm 1.05821  \hskip 0.5cm 0.97571 \hskip 0.5cm 0.92570   \hskip 0.5cm 0.90878 \\
 \mathsf{-1}\hskip 0.5cm 0.52576\hskip 0.5cm  0.43733 \hskip 0.4cm 0.38095  \hskip 0.3cm 0.34452   \hskip 0.3cm   0.32110\hskip 0.5cm   0.31276\hskip 0.7cm  1.57364 \hskip 0.5cm 1.26951   \hskip 0.5cm  1.04417   \hskip 0.5cm 0.95685  \hskip 0.5cm  0.91306    \hskip 0.5cm  0.89573  \\
 \mathsf{-0.5}\hskip 0.25cm 0.52151 \hskip 0.5cm  0.44530  \hskip 0.4cm  0.39563   \hskip 0.3cm   0.36700    \hskip 0.3cm  0.34674 \hskip 0.5cm 0.33839  \hskip 0.73cm  1.56899   \hskip 0.5cm 1.26603   \hskip 0.5cm  1.02228  \hskip 0.5cm   0.91709 \hskip 0.5cm  0.88295  \hskip 0.5cm  0.86733 \\
 \mathsf{0.0}\hskip 0.5cm  0.50074 \hskip 0.5cm  0.43772  \hskip 0.4cm   0.39748  \hskip 0.3cm 0.37900    \hskip 0.3cm 0.36711  \hskip 0.5cm 0.36009  \hskip 0.73cm 1.55840  \hskip 0.5cm 1.25445   \hskip 0.5cm  0.98989   \hskip 0.5cm 0.84925  \hskip 0.5cm 0.81847   \hskip 0.5cm 0.80948  \\
\end{array}
 $$
\end{table*}
%******************************************************************************%

%%%%%%%%%%%%%%%%%%%%%%%%%%% beta = 2, gamma = 5/3 %%%%%%%%%%%%%%%%%%%%%%%%%%%%%
\begin{table*} 
 \caption{Growth rates ($\delta_{r}$) for several  modes and   Mach number $M$, for $\gamma = 5/3$, $\beta = 2$, R = 1.}
 \label{g53b2r} %the labelling is gammabeta i
$$
\begin{array}{lccccccccc} 
\hline
\mathsf{ \,\,\,\,\,\,\,\,\,\,\,\,}
\hskip 3.5cm \rm fundamental \hskip 6.5cm \rm first \hskip 0.1cm  overtone  \\
\hline
\noalign{\smallskip} \noalign{\smallskip}
	 \mathsf{\alpha \,\,\,\,\,\,\,}
	 \hskip 0.6cm \mathsf{M = 1.4}
	 \hskip 0.6cm \mathsf{M = 2} 
	 \hskip 0.6cm \mathsf{M = 3}
         \hskip 0.6cm \mathsf{M = 5} 
	 \hskip 0.6cm \mathsf{M = 10}
         \hskip 0.5cm \mathsf{M = 40}
	 \hskip 0.9cm \mathsf{M = 1.4}
         \hskip 0.65cm \mathsf{M = 2} 
	 \hskip 0.65cm \mathsf{M = 3}
         \hskip 0.65cm \mathsf{M = 5} 
	 \hskip 0.6cm \mathsf{M = 10}
        \hskip 0.4cm \mathsf{M = 40}

    \\
 \hline  
 \noalign{\smallskip}
\mathsf{-2}\hskip 0.83cm 0.13189  \hskip 0.5cm  0.19818 \hskip 0.42cm  0.19427 \hskip 0.44cm  0.17816   \hskip 0.48cm  0.16996 \hskip 0.5cm  0.16729   \hskip 0.55cm  -0.05599   \hskip 0.46cm 0.26403   \hskip 0.47cm 0.36978  \hskip 0.5cm 0.35914 \hskip 0.47cm 0.34838    \hskip 0.5cm 0.34456 \\
 \mathsf{-1}\hskip 0.5cm -0.00441 \hskip 0.47cm 0.08220  \hskip 0.43cm 0.10656 \hskip 0.46cm 0.10601  \hskip 0.48cm 0.10393 \hskip 0.5cm 0.10314   \hskip 0.52cm -0.23776  \hskip 0.45cm 0.07674   \hskip 0.5cm 0.23030    \hskip 0.5cm  0.24461  \hskip 0.5cm  0.24104    \hskip 0.5cm  0.23943  \\ 
\mathsf{-0.5}\hskip 0.25cm -0.07480  \hskip 0.48cm 0.02392   \hskip 0.43cm 0.06172    \hskip 0.44cm 0.06845      \hskip 0.5cm 0.06920 \hskip 0.5cm 0.06929   \hskip 0.53cm   -0.33698   \hskip 0.1cm -0.02827  \hskip 0.48cm 0.14980  \hskip 0.52cm  0.18295  \hskip 0.5cm  0.18330  \hskip 0.5cm   0.18277 \\
 \mathsf{0.0}\hskip 0.5cm -0.14797  \hskip 0.1cm -0.03902   \hskip 0.42cm 0.01365   \hskip 0.46cm 0.02775    \hskip 0.5cm 0.03128   \hskip 0.5cm 0.03225 \hskip 0.53cm  -0.44318 \hskip 0.1cm  -0.14382    \hskip 0.5cm 0.05429  \hskip 0.5cm   0.11407 \hskip 0.5cm 0.11948  \hskip 0.52cm 0.12011  \\
\end{array}
 $$
\end{table*}
%******************************************************************************%
%%%%%%%%%%%%%%%%%%%%%%%EIGENFREQUENCIES TABLE%%%%%%%%%%%%%%%%%%%%%%%%%%%%%%%%%%
\begin{table*} 
 \caption{Eigenfrequencies ($\delta_{i}$) for several  modes and   Mach number $M$, for $\gamma = 5/3$, $\beta = 2$, R = 1. }
 \label{g53b2f} %the labelling is gammabeta i
$$
\begin{array}{lccccccccc} 
\hline
\mathsf{ \,\,\,\,\,\,\,\,\,\,\,\,}
\hskip 3.5cm \rm fundamental \hskip 6.5cm \rm first \hskip 0.1cm  overtone  \\
\hline
\noalign{\smallskip} \noalign{\smallskip}
	 \mathsf{\alpha \,\,\,\,\,\,\,}
	 \hskip 0.3cm \mathsf{M = 1.4}
	 \hskip 0.5cm \mathsf{M = 2} 
	 \hskip 0.5cm \mathsf{M = 3}
         \hskip 0.5cm \mathsf{M = 5} 
	 \hskip 0.5cm \mathsf{M = 10}
         \hskip 0.5cm \mathsf{M = 40}
	 \hskip 0.7cm \mathsf{M = 1.4}
         \hskip 0.65cm \mathsf{M = 2} 
	 \hskip 0.65cm \mathsf{M = 3}
         \hskip 0.5cm \mathsf{M = 5} 
	 \hskip 0.6cm \mathsf{M = 10}
        \hskip 0.5cm \mathsf{M = 40}
    \\
 \hline  
 \noalign{\smallskip}
\mathsf{-2}\hskip 0.5cm 0.49898   \hskip 0.5cm 0.39376 \hskip 0.5cm   0.30489   \hskip 0.4cm 0.24931 \hskip 0.3cm  0.22545  \hskip 0.3cm  0.21802  \hskip 0.7cm 1.50492 \hskip 0.5cm  1.27215  \hskip 0.5cm 1.14203 \hskip 0.49cm  1.03289 \hskip 0.35cm 0.97737 \hskip 0.5cm 0.95930 \\
\mathsf{-1}  \hskip 0.5cm 0.51095 \hskip 0.5cm 0.42256 \hskip 0.5cm  0.34659  \hskip 0.4cm  0.29390 \hskip 0.3cm  0.26972  \hskip 0.3cm   0.26210  \hskip 0.7cm  1.51568 \hskip 0.5cm 1.27464  \hskip 0.5cm  1.12782   \hskip 0.49cm 1.02682   \hskip 0.35cm 0.97073  \hskip 0.5cm  0.95221    \\
 \mathsf{-0.5}\hskip 0.27cm  0.50429  \hskip 0.5cm  0.42670   \hskip 0.5cm  0.35989  \hskip 0.38cm   0.31075     \hskip 0.3cm 0.28687 \hskip 0.3cm 0.27922   \hskip 0.73cm 1.51588    \hskip 0.48cm  1.26981 \hskip 0.5cm  1.10822  \hskip 0.5cm 1.01444  \hskip 0.35cm  0.95975   \hskip 0.48cm 0.94120  \\
 \mathsf{0.0}  \hskip 0.5cm 0.48748  \hskip 0.5cm  0.42303  \hskip 0.5cm 0.36770    \hskip 0.4cm 0.32441   \hskip 0.3cm 0.30149    \hskip 0.3cm 0.29389  \hskip 0.5cm \hskip 0.2cm 1.51247 \hskip 0.5cm  1.26093  \hskip 0.5cm 1.07935     \hskip 0.5cm  0.99119  \hskip 0.35cm 0.94052   \hskip 0.5cm  0.92221     \\
\end{array}
 $$
\end{table*}
%******************************************************************************%

%%%%%%%%%%%%%%%%%%%%%%%%%%% beta =1, gamma = 7/5 &&&&&&&&&&&&&&&&&&&&&&&&&&&&%%
\begin{table*} 
 \caption{Growth rates ($\delta_{r}$) for several  modes and   Mach number $M$, for $\gamma = 7/5$, $\beta = 1$, R = 1. }
 \label{g75b1r} %the labelling is gammabeta r
$$
\begin{array}{lccccccccc} 
\hline
\mathsf{ \,\,\,\,\,\,\,\,\,\,\,\,}
\hskip 4.cm \rm fundamental \hskip 6.5cm \rm first \hskip 0.1cm  overtone  \\
\hline
\noalign{\smallskip} \noalign{\smallskip}
	 \mathsf{\alpha \,\,\,\,\,\,\,}
	 \hskip 0.5cm \mathsf{M = 1.4}
	 \hskip 0.55cm \mathsf{M = 2} 
	 \hskip 0.6cm \mathsf{M = 3}
         \hskip 0.7cm \mathsf{M = 5} 
	 \hskip 0.6cm \mathsf{M = 10}
         \hskip 0.6cm \mathsf{M = 40}
	 \hskip 0.7cm \mathsf{M = 1.4}
         \hskip 0.6cm \mathsf{M = 2} 
	 \hskip 0.65cm \mathsf{M = 3}
         \hskip 0.65cm \mathsf{M = 5} 
	 \hskip 0.6cm \mathsf{M = 10}
         \hskip 0.45cm \mathsf{M = 40}
    \\
 \hline  
 \noalign{\smallskip}
\mathsf{-2} \hskip 0.3cm -0.08907 \hskip 0.55cm 0.03785   \hskip 0.5cm 0.09425  \hskip 0.5cm 0.08660    \hskip 0.5cm 0.07616   \hskip 0.5cm 0.07278    \hskip 0.35cm -0.43497  \hskip 0.2cm -0.12291 \hskip 0.57cm 0.12198  \hskip 0.58cm 0.20607  \hskip 0.42cm 0.19967 \hskip 0.43cm 0.19548  \\
 \mathsf{-1}\hskip 0.3cm -0.18752 \hskip 0.2cm -0.05832  \hskip 0.47cm  0.01949 \hskip 0.5cm  0.03530 \hskip 0.5cm 0.02885  \hskip 0.5cm 0.02726   \hskip 0.37cm -0.57272  \hskip 0.22cm -0.28452    \hskip 0.2cm -0.03932    \hskip 0.58cm 0.10080 \hskip 0.41cm  0.11901     \hskip 0.4cm   0.11778 \\
 \mathsf{-0.5} \hskip 0.1cm -0.24246  \hskip 0.15cm   -0.11341   \hskip 0.14cm -0.02624   \hskip 0.5cm 0.00082 \hskip 0.13cm -0.00116  \hskip 0.17cm -0.00203   \hskip 0.35cm -0.64925    \hskip 0.2cm  -0.37618  \hskip 0.2cm -0.13440  \hskip 0.58cm 0.02407  \hskip 0.4cm  0.06848 \hskip 0.42cm  0.06969  \\ 
 \mathsf{0.0}\hskip 0.3cm -0.30305 \hskip 0.2cm -0.17573 \hskip 0.1cm -0.08134   \hskip 0.14cm -0.04077    \hskip 0.13cm -0.03946   \hskip 0.17cm -0.04810  \hskip 0.35cm  -0.73312    \hskip 0.23cm -0.47828  \hskip 0.2cm  -0.24124 \hskip 0.2cm -0.07681 \hskip 0.4cm 0.00236  \hskip 0.42cm 0.01003\\
\end{array}
 $$
\end{table*}
%******************************************************************************%
%%%%%%%%%%%%%%%%%%%%%%%EIGENFREQUENCIES TABLE%%%%%%%%%%%%%%%%%%%%%%%%%%%%%%%%%%
\begin{table*} 
 \caption{Eigenfrequencies ($\delta_{i}$) for several  modes and   Mach number $M$, for $\gamma = 7/5$, $\beta = 1$, R = 1. }
 \label{g75b1f} %the labelling is gammabeta i
$$
\begin{array}{lccccccccc} 
\hline
\mathsf{ \,\,\,\,\,\,\,\,\,\,\,\,}
\hskip 3.5cm \rm fundamental \hskip 6.5cm \rm first \hskip 0.1cm  overtone  \\
\hline
\noalign{\smallskip} \noalign{\smallskip}
	 \mathsf{\alpha \,\,\,\,\,\,\,}
	 \hskip 0.3cm \mathsf{M = 1.4}
	 \hskip 0.5cm \mathsf{M = 2} 
	 \hskip 0.5cm \mathsf{M = 3}
         \hskip 0.5cm \mathsf{M = 5} 
	 \hskip 0.5cm \mathsf{M = 10}
         \hskip 0.5cm \mathsf{M = 40}
	 \hskip 0.7cm \mathsf{M = 1.4}
         \hskip 0.65cm \mathsf{M = 2} 
	 \hskip 0.65cm \mathsf{M = 3}
         \hskip 0.5cm \mathsf{M = 5} 
	 \hskip 0.7cm \mathsf{M = 10}
        \hskip 0.5cm \mathsf{M = 40}

    \\
 \hline  
 \noalign{\smallskip}
\mathsf{-2}\hskip 0.5cm 0.39767  \hskip 0.5cm   0.34552 \hskip 0.4cm  0.29282 \hskip 0.3cm 0.24754    \hskip 0.3cm 0.21898 \hskip 0.5cm  0.20952 \hskip 0.7cm 1.42551  \hskip 0.5cm 1.13970  \hskip 0.5cm 0.87852   \hskip 0.5cm  0.76775 \hskip 0.5cm 0.70466    \hskip 0.5cm  0.68182  \\
 \mathsf{-1}\hskip 0.5cm 0.36781 \hskip 0.52cm  0.33964  \hskip 0.4cm 0.30142   \hskip 0.3cm   0.26912    \hskip 0.3cm 0.24299  \hskip 0.5cm 0.23295     \hskip 0.7cm 1.42305    \hskip 0.5cm  1.14322  \hskip 0.5cm 0.86075  \hskip 0.5cm  0.72474  \hskip 0.5cm  0.67767  \hskip 0.5cm  0.65645   \\ 
 \mathsf{-0.5}\hskip 0.25cm 0.33927 \hskip 0.5cm 0.32565  \hskip 0.4cm  0.29805 \hskip 0.33cm 0.27453   \hskip 0.3cm 0.25310 \hskip 0.5cm 0.24316   \hskip 0.7cm  1.41915    \hskip 0.5cm  1.14098 \hskip 0.5cm   0.84834  \hskip 0.5cm 0.68781  \hskip 0.5cm  0.64858    \hskip 0.5cm 0.63132  \\
 \mathsf{0.0}\hskip 0.5cm  0.29464  \hskip 0.5cm 0.29888   \hskip 0.4cm  0.28562   \hskip 0.3cm  0.27146   \hskip 0.3cm 0.25888   \hskip 0.5cm 0.24996 \hskip 0.73cm 1.41348 \hskip 0.5cm  1.13617  \hskip 0.5cm 0.83129   \hskip 0.5cm  0.64222  \hskip 0.5cm  0.59451   \hskip 0.5cm   0.58847 \\
\end{array}
 $$
\end{table*}
%******************************************************************************%

%**************************beta = 2, gamma = 7/5*********************************
\begin{table*} 
 \caption{Growth rates ($\delta_{r}$) for several  modes and   Mach number $M$, for $\gamma = 7/5$, $\beta = 2$, R = 1. }
 \label{g75b2r} %the labelling is gammabeta r
$$
\begin{array}{lccccccccc} 
\hline
\mathsf{ \,\,\,\,\,\,\,\,\,\,\,\,}
\hskip 4.cm \rm fundamental \hskip 6.5cm \rm first \hskip 0.1cm  overtone  \\
\hline
\noalign{\smallskip} \noalign{\smallskip}
	 \mathsf{\alpha \,\,\,\,\,\,\,}
	 \hskip 0.5cm \mathsf{M = 1.4}
	 \hskip 0.55cm \mathsf{M = 2} 
	 \hskip 0.6cm \mathsf{M = 3}
         \hskip 0.7cm \mathsf{M = 5} 
	 \hskip 0.6cm \mathsf{M = 10}
         \hskip 0.6cm \mathsf{M = 40}
	 \hskip 0.7cm \mathsf{M = 1.4}
         \hskip 0.6cm \mathsf{M = 2} 
	 \hskip 0.65cm \mathsf{M = 3}
         \hskip 0.65cm \mathsf{M = 5} 
	 \hskip 0.6cm \mathsf{M = 10}
         \hskip 0.45cm \mathsf{M = 40}
    \\
 \hline  
 \noalign{\smallskip}
\mathsf{-2} \hskip 0.32cm -0.09916 \hskip 0.52cm 0.02373  \hskip 0.47cm 0.06491  \hskip 0.45cm 0.06548   \hskip 0.5cm 0.06276  \hskip 0.5cm  0.06168 \hskip 0.4cm   -0.38508 \hskip 0.2cm -0.02766    \hskip 0.54cm 0.18108 \hskip 0.57cm  0.20098 \hskip 0.4cm  0.19312\hskip 0.49cm 0.18986    \\
 \mathsf{-1}\hskip 0.3cm -0.18016 \hskip 0.2cm -0.04894 \hskip 0.45cm 0.01531 \hskip 0.45cm 0.02769   \hskip 0.5cm 0.02924     \hskip 0.5cm 0.02951   \hskip 0.38cm -0.50806  \hskip 0.2cm -0.16818 \hskip 0.55cm 0.07156 \hskip 0.57cm 0.13468  \hskip 0.43cm 0.13438  \hskip 0.47cm 0.13306     \\ 
 \mathsf{-0.5} \hskip 0.07cm -0.22339   \hskip 0.17cm -0.08837   \hskip 0.1cm -0.01312  \hskip 0.48cm 0.00598  \hskip 0.5cm 0.00976  \hskip 0.49cm 0.01074   \hskip 0.4cm -0.57519   \hskip 0.2cm  -0.24699 \hskip 0.54cm 0.00248   \hskip 0.57cm 0.09458  \hskip 0.45cm  0.10022  \hskip 0.45cm 0.09991  \\ 
 \mathsf{0.0}\hskip 0.3cm -0.26895 \hskip 0.2cm -0.13126  \hskip 0.1cm -0.04541  \hskip 0.13cm -0.01860    \hskip 0.1cm -0.01207   \hskip 0.15cm -0.01087  \hskip 0.38cm -0.64711    \hskip 0.2cm -0.33298  \hskip 0.2cm -0.07822   \hskip 0.53cm 0.04517  \hskip 0.45cm 0.06099  \hskip 0.45cm  0.06204\\
\end{array}
 $$
\end{table*}
%******************************************************************************%%%%%%%%%%%%%%%%

%%%%%%%%%%%%%%%%%%%%%%%EIGENFREQUENCIES TABLE%%%%%%%%%%%%%%%%%%%%%%%%%%%%%%%%%%%%%%%%%%%%%%%%%%%
\begin{table*} 
 \caption{Eigenfrequencies ($\delta_{i}$) for several  modes and   Mach number $M$, for $\gamma = 7/5$, $\beta = 2$, R = 1. }
 \label{g75b2f} %the labelling is gammabeta i
$$
\begin{array}{lccccccccc} 
\hline
\mathsf{ \,\,\,\,\,\,\,\,\,\,\,\,}
\hskip 3.5cm \rm fundamental \hskip 6.5cm \rm first \hskip 0.1cm  overtone  \\
\hline
\noalign{\smallskip} \noalign{\smallskip}
	 \mathsf{\alpha \,\,\,\,\,\,\,}
	 \hskip 0.3cm \mathsf{M = 1.4}
	 \hskip 0.5cm \mathsf{M = 2} 
	 \hskip 0.5cm \mathsf{M = 3}
         \hskip 0.5cm \mathsf{M = 5} 
	 \hskip 0.5cm \mathsf{M = 10}
         \hskip 0.5cm \mathsf{M = 40}
	 \hskip 0.7cm \mathsf{M = 1.4}
         \hskip 0.65cm \mathsf{M = 2} 
	 \hskip 0.65cm \mathsf{M = 3}
         \hskip 0.5cm \mathsf{M = 5} 
	 \hskip 0.7cm \mathsf{M = 10}
        \hskip 0.5cm \mathsf{M = 40}
    \\
 \hline  
 \noalign{\smallskip}
\mathsf{-2}\hskip 0.5cm 0.38868 \hskip 0.5cm 0.33930 \hskip 0.4cm 0.27263 \hskip 0.3cm 0.21689    \hskip 0.3cm  0.19035  \hskip 0.5cm  0.18196 \hskip 0.7cm   1.38127 \hskip 0.5cm 1.13340  \hskip 0.5cm  0.93707 \hskip 0.47cm 0.80891 \hskip 0.49cm  0.73527   \hskip 0.5cm  0.71091 \\
 \mathsf{-1}\hskip 0.5cm 0.35902 \hskip 0.52cm 0.33269  \hskip 0.37cm  0.28334  \hskip 0.3cm  0.23446    \hskip 0.3cm  0.20783  \hskip 0.5cm   0.19917   \hskip 0.72cm  1.38221   \hskip 0.5cm 1.13281   \hskip 0.5cm  0.90820 \hskip 0.45cm 0.79096  \hskip 0.5cm 0.72107  \hskip 0.5cm 0.69689   \\ 
 \mathsf{-0.5} \hskip 0.3cm 0.33488  \hskip 0.47cm  0.32298   \hskip 0.4cm 0.28461  \hskip 0.3cm 0.24175 \hskip 0.3cm  0.21574  \hskip 0.5cm  0.20699    \hskip 0.7cm 1.38113   \hskip 0.5cm 1.13056   \hskip 0.5cm  0.89052  \hskip 0.46cm  0.77352  \hskip 0.5cm  0.70869   \hskip 0.5cm 0.68488   \\ 
\mathsf{0.0}  \hskip 0.5cm 0.30130  \hskip 0.5cm  0.30690 \hskip 0.4cm 0.28163  \hskip 0.3cm 0.24715  \hskip 0.3cm  0.22278   \hskip 0.5cm  0.21404   \hskip 0.746cm 1.37917  \hskip 0.46cm  1.12659    \hskip 0.5cm0.87115     \hskip 0.47cm 0.74621 \hskip 0.5cm 0.68991 \hskip 0.5cm   0.66708  \\
\end{array}
 $$
\end{table*}
%******************************************************************************%&&&%%&&&%%&&&%%

%%%%%%%%%%%%%%%%%%%%%%%%%%% beta = 1, gamma = 10/7 &&&&&&&&&&&&&&&&&&&&&&&&&&&&%%&&&%%&&&%%&&&%%
\begin{table*}
 \caption{Growth rates ($\delta_{r}$) for several modes and Mach number $M$, for $\gamma = 10/7$, $\beta = 1$, R = 1. }
 \label{g107b1r} %the labelling is gammabeta r
$$
\begin{array}{lccccccccc} 
\hline
\mathsf{ \,\,\,\,\,\,\,\,\,\,\,\,}
\hskip 4.cm \rm fundamental \hskip 6.5cm \rm first \hskip 0.1cm  overtone  \\
\hline
\noalign{\smallskip} \noalign{\smallskip}
	 \mathsf{\alpha \,\,\,\,\,\,\,}
	 \hskip 0.5cm \mathsf{M = 1.4}
	 \hskip 0.55cm \mathsf{M = 2} 
	 \hskip 0.6cm \mathsf{M = 3}
         \hskip 0.7cm \mathsf{M = 5} 
	 \hskip 0.6cm \mathsf{M = 10}
         \hskip 0.6cm \mathsf{M = 40}
	 \hskip 0.7cm \mathsf{M = 1.4}
         \hskip 0.6cm \mathsf{M = 2} 
	 \hskip 0.65cm \mathsf{M = 3}
         \hskip 0.65cm \mathsf{M = 5} 
	 \hskip 0.6cm \mathsf{M = 10}
         \hskip 0.45cm \mathsf{M = 40}
    \\
 \hline  
 \noalign{\smallskip}
\mathsf{-2} \hskip 0.3cm -0.06194  \hskip 0.55cm 0.06121   \hskip 0.5cm 0.11186  \hskip 0.53cm   0.10098   \hskip 0.5cm 0.08948    \hskip 0.5cm  0.08577   \hskip 0.35cm-0.39813   \hskip 0.2cm -0.08906  \hskip 0.54cm 0.14898  \hskip 0.58cm 0.22356  \hskip 0.42cm 0.21610  \hskip 0.43cm 0.21184 \\
 \mathsf{-1}\hskip 0.3cm -0.16697 \hskip 0.2cm -0.04057  \hskip 0.47cm 0.03250   \hskip 0.53cm 0.04351   \hskip 0.5cm 0.03815  \hskip 0.5cm  0.03632  \hskip 0.35cm  -0.54360  \hskip 0.2cm  -0.25787    \hskip 0.2cm -0.01852    \hskip 0.58cm 0.11383 \hskip 0.42cm 0.12907      \hskip 0.4cm    0.12777\\
 \mathsf{-0.5} \hskip 0.1cm -0.22552  \hskip 0.15cm -0.09868     \hskip 0.14cm -0.01555   \hskip 0.53cm 0.00856 \hskip 0.5cm 0.00596  \hskip 0.5cm 0.00489  \hskip 0.35cm -0.62430    \hskip 0.2cm  -0.35359  \hskip 0.2cm -0.11778   \hskip 0.58cm 0.03503  \hskip 0.4cm 0.07514 \hskip 0.42cm  0.07619  \\ 
 \mathsf{0.0}\hskip 0.3cm  -0.29004 \hskip 0.2cm -0.16437 \hskip 0.1cm -0.07341    \hskip 0.16cm -0.03552    \hskip 0.13cm -0.03524  \hskip 0.17cm -0.03652  \hskip 0.35cm  -0.71257    \hskip 0.23cm -0.45997  \hskip 0.2cm  -0.22948  \hskip 0.2cm -0.06953 \hskip 0.4cm 0.00551  \hskip 0.42cm 0.01260 \\
\end{array}
 $$
\end{table*}
%******************************************************************************%%%%%%%%%%%%%%

%%%%%%%%%%%%%%%%%%%%%%%EIGENFREQUENCIES TABLE%%%%%%%%%%%%%%%%%%%%%%%%%%%%%%%%%%%%%%%%%%%%%%
\begin{table*}
 \caption{Eigenfrequencies ($\delta_{i}$) for several  modes and   Mach number $M$, for $\gamma = 10/7$, $\beta = 1$, R = 1. }
 \label{g107b1f} %the labelling is gammabeta i
$$
\begin{array}{lccccccccc} 
\hline
\mathsf{ \,\,\,\,\,\,\,\,\,\,\,\,}
\hskip 3.5cm \rm fundamental \hskip 6.5cm \rm first \hskip 0.1cm  overtone  \\
\hline
\noalign{\smallskip} \noalign{\smallskip}
	 \mathsf{\alpha \,\,\,\,\,\,\,}
	 \hskip 0.3cm \mathsf{M = 1.4}
	 \hskip 0.5cm \mathsf{M = 2} 
	 \hskip 0.5cm \mathsf{M = 3}
         \hskip 0.5cm \mathsf{M = 5} 
	 \hskip 0.5cm \mathsf{M = 10}
         \hskip 0.5cm \mathsf{M = 40}
	 \hskip 0.7cm \mathsf{M = 1.4}
         \hskip 0.65cm \mathsf{M = 2} 
	 \hskip 0.65cm \mathsf{M = 3}
         \hskip 0.5cm \mathsf{M = 5} 
	 \hskip 0.7cm \mathsf{M = 10}
        \hskip 0.5cm \mathsf{M = 40}
    \\
 \hline  
 \noalign{\smallskip}
\mathsf{-2}\hskip 0.5cm 0.41642   \hskip 0.5cm 0.35499   \hskip 0.4cm 0.29967  \hskip 0.3cm  0.25395   \hskip 0.3cm 0.22603  \hskip 0.5cm 0.21684   \hskip 0.7cm  1.44303 \hskip 0.5cm 1.15321  \hskip 0.5cm 0.90101   \hskip 0.5cm 0.79515 \hskip 0.5cm 0.73329   \hskip 0.5cm 0.71121   \\
 \mathsf{-1}\hskip 0.5cm  0.39227 \hskip 0.52cm  0.35418   \hskip 0.4cm 0.31234    \hskip 0.3cm  0.27901     \hskip 0.3cm 0.25313  \hskip 0.5cm 0.24332   \hskip 0.7cm  1.44112   \hskip 0.5cm  1.15748   \hskip 0.5cm 0.88180  \hskip 0.5cm 0.75349   \hskip 0.5cm 0.70664   \hskip 0.5cm 0.68589    \\ 
 \mathsf{-0.5}\hskip 0.25cm 0.36680 \hskip 0.5cm  0.34254  \hskip 0.4cm 0.31064  \hskip 0.33cm 0.28603  \hskip 0.3cm 0.26464 \hskip 0.5cm  0.25492 \hskip 0.7cm 1.43701     \hskip 0.5cm 1.15503   \hskip 0.5cm 0.86817    \hskip 0.5cm 0.71530   \hskip 0.5cm 0.67711    \hskip 0.5cm  0.65994  \\
 \mathsf{0.0}\hskip 0.5cm  0.32614   \hskip 0.5cm  0.31826  \hskip 0.4cm 0.29969    \hskip 0.3cm 0.28448   \hskip 0.3cm 0.27185   \hskip 0.5cm    0.26320 \hskip 0.73cm  1.43075  \hskip 0.5cm 1.14950  \hskip 0.5cm  0.84948   \hskip 0.5cm 0.66645  \hskip 0.5cm 0.62159   \hskip 0.5cm  0.61500  \\
\end{array}
 $$
\end{table*}
%******************************************************************************%

%**************************beta = 2, gamma = 10/7*********************************
\begin{table*}
 \caption{Growth rates ($\delta_{r}$) for several modes and Mach number $M$, for $\gamma = 10/7$, $\beta = 2$, R = 1. }
 \label{g107b2r} %the labelling is gammabeta r
$$
\begin{array}{lccccccccc} 
\hline
\mathsf{ \,\,\,\,\,\,\,\,\,\,\,\,}
\hskip 4.cm \rm fundamental \hskip 6.5cm \rm first \hskip 0.1cm  overtone  \\
\hline
\noalign{\smallskip} \noalign{\smallskip}
	 \mathsf{\alpha \,\,\,\,\,\,\,}
	 \hskip 0.5cm \mathsf{M = 1.4}
	 \hskip 0.55cm \mathsf{M = 2} 
	 \hskip 0.6cm \mathsf{M = 3}
         \hskip 0.7cm \mathsf{M = 5} 
	 \hskip 0.6cm \mathsf{M = 10}
         \hskip 0.6cm \mathsf{M = 40}
	 \hskip 0.7cm \mathsf{M = 1.4}
         \hskip 0.6cm \mathsf{M = 2} 
	 \hskip 0.65cm \mathsf{M = 3}
         \hskip 0.65cm \mathsf{M = 5} 
	 \hskip 0.6cm \mathsf{M = 10}
         \hskip 0.45cm \mathsf{M = 40}
    \\
 \hline  
 \noalign{\smallskip}
\mathsf{-2} \hskip 0.32cm  -0.07375  \hskip 0.52cm 0.04142  \hskip 0.5cm 0.07763   \hskip 0.45cm 0.07614   \hskip 0.5cm 0.07280  \hskip 0.5cm 0.07155  \hskip 0.4cm  -0.34620  \hskip 0.52cm  0.08304   \hskip 0.54cm 0.20306  \hskip 0.57cm 0.21684   \hskip 0.4cm 0.20862 \hskip 0.53cm  0.20533   \\
 \mathsf{-1}\hskip 0.3cm -0.16013  \hskip 0.2cm  -0.03363 \hskip 0.5cm 0.02499   \hskip 0.45cm 0.03545   \hskip 0.5cm  0.03653  \hskip 0.5cm  0.03665  \hskip 0.38cm  -0.47564   \hskip 0.2cm -0.13749 \hskip 0.54cm 0.09213  \hskip 0.57cm 0.14628  \hskip 0.4cm 0.14536  \hskip 0.53cm 0.14402     \\ 
 \mathsf{-0.5} \hskip 0.1cm  -0.20616 \hskip 0.2cm -0.07494   \hskip 0.1cm  -0.00486 \hskip 0.45cm 0.01226  \hskip 0.5cm 0.01562  \hskip 0.5cm  0.01647  \hskip 0.4cm -0.54633   \hskip 0.2cm  -0.21941   \hskip 0.54cm 0.02188   \hskip 0.55cm 0.10428  \hskip 0.4cm 0.10882   \hskip 0.53cm 0.10856   \\ 
 \mathsf{0.0}\hskip 0.3cm -0.25462  \hskip 0.2cm -0.11983   \hskip 0.14cm -0.03857  \hskip 0.13cm  -0.01390   \hskip 0.1cm  -0.00827 \hskip 0.15cm  -0.00666 \hskip 0.38cm   -0.62207    \hskip 0.2cm -0.30888  \hskip 0.2cm   -0.06088 \hskip 0.5cm 0.05333 \hskip 0.43cm  0.06702\hskip 0.53cm 0.06801  \\
\end{array}
 $$
\end{table*}
%******************************************************************************%

%%%%%%%%%%%%%%%%%%%%%%%EIGENFREQUENCIES TABLE%%%%%%%%%%%%%%%%%%%%%%%%%%%%%%%%%%
\begin{table*}
 \caption{Eigenfrequencies ($\delta_{i}$) for several  modes and   Mach number $M$, for $\gamma = 10/7$, $\beta = 2$, R = 1. }
 \label{g107b2f} %the labelling is gammabeta i
$$
\begin{array}{lccccccccc} 
\hline
\mathsf{ \,\,\,\,\,\,\,\,\,\,\,\,}
\hskip 3.5cm \rm fundamental \hskip 6.5cm \rm first \hskip 0.1cm  overtone  \\
\hline
\noalign{\smallskip} \noalign{\smallskip}
	 \mathsf{\alpha \,\,\,\,\,\,\,}
	 \hskip 0.3cm \mathsf{M = 1.4}
	 \hskip 0.5cm \mathsf{M = 2} 
	 \hskip 0.5cm \mathsf{M = 3}
         \hskip 0.5cm \mathsf{M = 5} 
	 \hskip 0.5cm \mathsf{M = 10}
         \hskip 0.5cm \mathsf{M = 40}
	 \hskip 0.7cm \mathsf{M = 1.4}
         \hskip 0.65cm \mathsf{M = 2} 
	 \hskip 0.65cm \mathsf{M = 3}
         \hskip 0.5cm \mathsf{M = 5} 
	 \hskip 0.7cm \mathsf{M = 10}
        \hskip 0.5cm \mathsf{M = 40}
    \\
 \hline  
 \noalign{\smallskip}
\mathsf{-2}\hskip 0.5cm  0.40777  \hskip 0.5cm 0.34899  \hskip 0.4cm 0.27836   \hskip 0.3cm 0.22238    \hskip 0.3cm 0.19627  \hskip 0.5cm 0.18805   \hskip 0.7cm 1.39652  \hskip 0.5cm 1.14917   \hskip 0.5cm 0.96382 \hskip 0.47cm 0.83731 \hskip 0.49cm  0.76595    \hskip 0.5cm  0.74242  \\
 \mathsf{-1}\hskip 0.5cm 0.38242  \hskip 0.52cm 0.34567  \hskip 0.37cm 0.29182   \hskip 0.3cm 0.24202  \hskip 0.3cm 0.21576  \hskip 0.5cm 0.20728 \hskip 0.72cm   1.39809  \hskip 0.5cm 1.14834  \hskip 0.5cm 0.93484 \hskip 0.45cm  0.82038   \hskip 0.5cm 0.75204  \hskip 0.5cm  0.72862   \\ 
 \mathsf{-0.5} \hskip 0.3cm 0.36056 \hskip 0.47cm  0.33752    \hskip 0.4cm  0.29438  \hskip 0.3cm 0.25021 \hskip 0.3cm 0.22449   \hskip 0.5cm  0.21592    \hskip 0.7cm 1.39700   \hskip 0.5cm 1.14576   \hskip 0.5cm  0.91599  \hskip 0.46cm  0.80349  \hskip 0.5cm  0.73960   \hskip 0.5cm 0.71648   \\ 
\mathsf{0.0}  \hskip 0.5cm 0.32969 \hskip 0.5cm 0.32302  \hskip 0.4cm 0.29263  \hskip 0.3cm 0.25649  \hskip 0.3cm  0.23227   \hskip 0.5cm  0.22371    \hskip 0.746cm  1.39478  \hskip 0.46cm 1.14127     \hskip 0.5cm 0.89488     \hskip 0.47cm  0.77648 \hskip 0.5cm 0.72056  \hskip 0.5cm 0.69829   \\
\end{array}
 $$
\end{table*}
%******************************************************************************%

%%%%%%%%%%%%%%%%%%%%%%%%%%%%%%%%%%%%%% Critical values of alpha for R =1%%%%%%%
\begin{table*}
 \caption{Critical values of $\alpha$ for various modes and Mach number M of different cases for R = 1. }
 \label{b12R1} %the labelling is gammabeta i
$$
\begin{array}{lcccccc} 
\hline
%\mathsf{ \,\,\,\,\,\,\,\,\,\,\,\,}
\hskip 2.5cm \rm fundamental \hskip 4.5cm \rm first \hskip 0.1cm  overtone \\
\hline
\noalign{\smallskip} 
	 \mathsf{(\beta, \gamma) }
	\hskip 0.65cm \mathsf{M = 1.4 }
	 \hskip 0.65cm \mathsf{M = 2} 
	 \hskip 0.65cm \mathsf{M = 3}
         \hskip 0.6cm \mathsf{M = 5} 
         \hskip 1.cm  \mathsf{M = 1.4 }
	 \hskip 0.65cm \mathsf{M = 2} 
	 \hskip 0.75cm \mathsf{M = 3}
         \hskip 0.75cm \mathsf{M = 5}
   \\
 \hline  
 \noalign{\smallskip}
\mathsf{(1,5/3)}\hskip 0.3cm -1.029  \hskip 0.5cm -0.423  \hskip 0.4cm -0.067  \hskip 0.66cm 0.081  \hskip 1.1cm -2.613 \hskip 0.53cm -1.280 \hskip 0.5cm -0.491 \hskip 0.57cm -0.058  \\
 \mathsf{(2,5/3)}\hskip 0.3cm -1.031 \hskip 0.5cm -0.309  \hskip 0.74cm 0.134 \hskip 0.665cm 0.312  \hskip 1.1cm -2.330  \hskip 0.53cm -0.630 \hskip 0.86cm 0.246 \hskip 0.95cm 0.659  \\ 
\mathsf{(1,7/5)}\hskip 0.3cm -3.009 \hskip 0.5cm  -1.588 \hskip 0.4cm -0.776 \hskip 0.3cm  -0.488 \hskip 1.1cm  -6.044 \hskip 0.5cm  -2.870  \hskip 0.5cm -1.222  \hskip 0.6cm -0.368    \\
 \mathsf{(2,7/5)}\hskip 0.3cm -3.320  \hskip 0.5cm  -1.673  \hskip 0.4cm -0.723   \hskip 0.3cm  -0.372  \hskip 1.1cm -5.898 \hskip 0.5cm  -2.214 \hskip 0.5cm -0.483  \hskip 0.95cm 0.359     \\ 
\mathsf{(1,10/7)}\hskip 0.15cm -2.646   \hskip 0.5cm -1.380  \hskip 0.4cm -0.652  \hskip 0.3cm -0.391 \hskip 1.1cm -5.421 \hskip 0.5cm -2.593  \hskip 0.5cm -1.100 \hskip 0.6cm -0.317    \\
 \mathsf{(2,10/7)}\hskip 0.15cm -2.907 \hskip 0.5cm  -1.434  \hskip 0.4cm -0.577   \hskip 0.3cm  -0.257 \hskip 1.1cm -5.255 \hskip 0.5cm -1.939 \hskip 0.5cm -0.359 \hskip 0.95cm 0.409\\
\end{array}
 $$
\end{table*}
%******************************************************************************%

%%%%%%%%%%%%%%%%%%%%%%%%%%%%%%%%%%%%%% Critical values of alpha for R =0%%%%%%%

\begin{table*}
 \caption{Critical values of $\alpha$ for various modes and Mach number M of different cases for R = 0. }
 \label{b12R0} %the labelling is gammabeta i
$$
\begin{array}{lcccccc} 
\hline
%\mathsf{ \,\,\,\,\,\,\,\,\,\,\,\,}
\hskip 2.5cm \rm fundamental \hskip 4.5cm \rm first \hskip 0.1cm  overtone \\
\hline
\noalign{\smallskip} 
	 \mathsf{(\beta, \gamma) }
	\hskip 0.65cm \mathsf{M = 1.4 }
	 \hskip 0.65cm \mathsf{M = 2} 
	 \hskip 0.65cm \mathsf{M = 3}
         \hskip 0.6cm \mathsf{M = 5} 
         \hskip 1.cm  \mathsf{M = 1.4 }
	 \hskip 0.65cm \mathsf{M = 2} 
	 \hskip 0.75cm \mathsf{M = 3}
         \hskip 0.75cm \mathsf{M = 5}
   \\
 \hline  
 \noalign{\smallskip}
\mathsf{(1,5/3)}\hskip 0.66cm  0.407  \hskip 0.86cm 0.201  \hskip 0.75cm 0.106  \hskip 0.66cm 0.068  \hskip 1.1cm  -0.106 \hskip 0.53cm -0.071\hskip 0.86cm 0.047 \hskip 0.9cm  0.114 \\
 \mathsf{(2,5/3)}\hskip 0.66cm 0.014 \hskip 0.86cm 0.002  \hskip 0.76cm 0.183 \hskip 0.665cm 0.310  \hskip 1.08cm  -0.019 \hskip 0.89cm 0.366\hskip 0.86cm 0.604 \hskip 0.9cm 0.719  \\ 
\mathsf{(1,7/5)}\hskip 0.66cm  0.220\hskip 0.5cm -0.200  \hskip 0.41cm -0.413 \hskip 0.3cm -0.495  \hskip 1.08cm -0.377  \hskip 0.5cm -0.336   \hskip 0.5cm  -0.098 \hskip 0.9cm  0.022   \\
 \mathsf{(2,7/5)}\hskip 0.3cm -0.794  \hskip 0.5cm -0.886   \hskip 0.4cm -0.585  \hskip 0.3cm -0.370 \hskip 1.08cm  -0.752\hskip 0.5cm -0.078 \hskip 0.88cm 0.349  \hskip 0.9cm  0.555    \\ 
\mathsf{(1,10/7)}\hskip 0.53cm 0.248  \hskip 0.49cm -0.133  \hskip 0.4cm  -0.325  \hskip 0.3cm -0.399 \hskip 1.08cm -0.333\hskip 0.5cm  -0.292 \hskip 0.5cm -0.076 \hskip 0.9cm 0.036   \\
 \mathsf{(2,10/7)}\hskip 0.15cm -0.660 \hskip 0.5cm  -0.737  \hskip 0.4cm  -0.456  \hskip 0.3cm -0.256 \hskip 1.1cm -0.631 \hskip 0.5cm  -0.006 \hskip 0.86cm  0.389\hskip 0.9cm  0.579\\
\end{array}
 $$
\end{table*}
%******************************************************************************%

\label{lastpage}
\end{document}